\documentclass[aps,pra,twocolumn,superscriptaddress,showpacs,amsmath,amssymb,floatfix]{revtex4}
\usepackage{graphicx}
\usepackage{hyperref}
\allowdisplaybreaks
\begin{document}

\title{Dipolar Bose-Einstein Condensates in Weak Anisotropic Disorder}

\author{Branko Nikoli\'c}
\email[]{branko.nikolic@fu-berlin.de}
\affiliation{Institut f\"ur Theoretische Physik, Freie Universit\"at Berlin, Arnimallee 14, 14195 Berlin, Germany}
\affiliation{Scientific Computing Laboratory, Institute of Physics Belgrade, University of Belgrade, Pregrevica 118, 11080 Belgrade, Serbia}

\author{Antun Bala\v z}
\email[]{antun@ipb.ac.rs}
\affiliation{Scientific Computing Laboratory, Institute of Physics Belgrade, University of Belgrade, Pregrevica 118, 11080 Belgrade, Serbia}
\affiliation{Hanse-Wissenschaftskolleg, Lehmkuhlenbusch 4, 27753 Delmenhorst, Germany}

\author{Axel Pelster}
\email[]{axel.pelster@physik.uni-kl.de}
\affiliation{Hanse-Wissenschaftskolleg, Lehmkuhlenbusch 4, 27753 Delmenhorst, Germany}
\affiliation{Fachbereich Physik und Forschungszentrum OPTIMAS, Technische
Universit\"at Kaiserslautern, Erwin-Schr\"odinger Stra{\ss}e, Geb\"aude 46, 67663 Kaiserslautern, Germany}

\date{\today}

\begin{abstract}
Here we study properties of a homogeneous dipolar Bose-Einstein condensate in a weak anisotropic random potential with Lorentzian correlation
at zero temperature.
To this end we solve perturbatively the Gross-Pitaevskii equation to second order in the random potential strength
and obtain analytic results for the disorder ensemble averages of both the condensate and the superfluid depletion, the equation of state, 
and the sound velocity.
For a pure contact interaction and a vanishing correlation length, we reproduce the seminal results of Huang
and Meng, which were originally derived within a Bogoliubov theory around a disorder-averaged
background field. For dipolar interaction and isotropic Lorentzian-correlated disorder, 
we obtain results which are qualitatively similar to the case of an isotropic Gaussian-correlated disorder. 
In the case of an anisotropic disorder,
the physical observables show characteristic anisotropies which arise from
the formation of fragmented dipolar condensates in the local minima of the disorder potential.
\end{abstract}


\maketitle

\section{Introduction}
Since the realization of Bose-Einstein condensates (BECs) in 1995 \cite{anderson1995,davis1995}, there was a significant interest about effects of the disordered potentials on the properties of ultracold quantum gases \cite{shapiro2012}. The reason for this is not only because of the unavoidable irregularities in the trapping potential induced by wire imperfections \cite{kruger2007,fortagh2007}, but also due to the fact that disorder can be generated and controlled using laser speckles \cite{dainty1984,billy2008}. It is well known that cold atoms are a promising tool for simulating other physical systems \cite{lewenstein2007} in the sense of Feynman's quantum simulator \cite{feynman1982}. This applies also to the phenomenon of Anderson localization, which was originally used to microscopically describe the absence of diffusion in terms of disorder \cite{anderson1958}. It has a clear BEC analogue \cite{paul2007}, which has been directly observed \cite{billy2008,deissler2010}. Also, localization inside BECs due to disorder created by atomic impurities on a lattice was studied theoretically \cite{gavish2005} and recently observed experimentally \cite{gadway2011}.

For a theoretical analysis of global dirty boson properties, different methods have been used to describe various limits, ranging from 
the Bogoliubov theory \cite{huang1992,giorgini1994,kobayashi2002,lugan2007,falco2007,hu2009,gaul2011,lugan2011,muller2012}, numerical approaches \cite{muruganandam2009,cai2010,min2012,vudragovic2012}, to the Parisi replica method \cite{lopatin2002,graham2008,morrison2008,bhongale2010,han2011}. It turns out that
long-range correlations within both the condensate and the superfluid remain, despite the presence of disorder. However, both quantities are depleted due to the localization of fragmented condensates in the local disorder potential minima. For a strong enough disorder in a homogeneous system, the depletion increases to such an extent that even a critical disorder strength exists, above which a Bose-glass phase appears, consisting only of localized mini-condensates \cite{navez2007,yukalov2007,nattermann2008,falco2009,falco2009a,graham2009}. Effects of disorder were also studied for harmonically trapped BECs \cite{timmer2006,falco2007a,nattermann2008,falco2009a} and BECs in optical lattices \cite{fisher1989,krutitsky2006,morrison2008,hu2009,bissbort2009,muller2012}, while the temperature behavior of dirty boson properties was examined in
Refs.~\cite{huang1992,lopatin2002,krutitsky2006,timmer2006,falco2007,graham2009,oliveira2010,han2011}.

Realization of atomic dipolar BECs \cite{griesmaier2005,lu2010,aikawa2012} with long-range anisotropic interaction has generated large interest in the theory of 
dipolar quantum gases 
\cite{goral2000,lahaye2009,cremon2010,muruganandam2010,lima2010r,lima2010,adhikari2012,lima2011r,lima2012}.
Increase in the strength of dipolar interaction is possible by substituting atoms with magnetic dipoles by heteronuclear molecules, which have a strong electric dipolar moment in rovibrational ground state \cite{ni2008}, or by inducing radiative coupling by placing dipoles into a resonator \cite{mottl2012}. Dipolar condensates were studied in the case of isotropic disorder \cite{falco2009,krumnow2011},
which yields characteristic anisotropies for both the superfluid density and the sound velocity at zero temperature due to the anisotropy of the dipolar interaction. 
Although a 3D isotropic laser speckle potential has recently been proposed in Ref.~\cite{Bakhodir},
the typical disorder realized in experiments is cylindrically symmetric and, to the best of our knowledge, it has so far been examined only
numerically for contact interaction \cite{pezze2011,piraud2012}.
Therefore, motivated by the experiments with dipolar BECs in anisotropic disorder potentials,  we develop in this paper a mean-field theory and analytically study the impact of a weak anisotropic disorder on physical properties of a polarized dipolar BEC at zero temperature.

To this end we proceed as follows. Following the approach developed in Ref.~\cite{krumnow2011}, in Sec.~\ref{sec:mf} we calculate the lowest-order corrections of BEC properties due to the presence of disorder within a mean-field theory. For the sake of generality we consider an arbitrary two-particle interaction and a general disorder correlation function. In Sec.~\ref{sec:diplor} we specialize the developed formalism to dipolar interaction and a Lorentzian-correlated disorder in Fourier space. This yields for both the superfluid density and the sound velocity characteristic anisotropies, which should be measurable in an experiment. In Sec.~\ref{sec:con} we present our conclusions and outlook for further related research. Finally, the Appendix gives analytical results for the condensate depletion and the disorder correction to the chemical potential, while the Supplemental Material \cite{supplemental} gives complete analytical results for the disorder correction of both the superfluid density and the sound velocity for the general case of a BEC with contact and dipole-dipole interaction in an anisotropic, Lorentzian-correlated disorder.

\section{Mean-field approach for weak disorder}
\label{sec:mf}
Bogoliubov quasiparticles and disorder induced fluctuations decouple in the lowest order
\cite{huang1992,giorgini1994,kobayashi2002,lugan2007,falco2007,hu2009,gaul2011,lugan2011,muller2012},
suggesting that disorder corrections can be calculated at zero temperature by neglecting quantum fluctuations and using a mean-field macroscopic wave function $\psi(\mathbf{r})$ governed by the time-independent Gross-Pitaevskii (GP) equation:
\begin{eqnarray}
&&\hspace*{-1cm}
\bigg(-\frac{\hbar^2}{2m}\Delta+\int d^3 r'\, V(\mathbf{r}-\mathbf{r}')
\psi^*(\mathbf{r'})\psi(\mathbf{r'}) \nonumber\\
&&\hspace*{1cm}+ U(\mathbf{r})-\mu\bigg)\psi(\mathbf{r})=0\, .
\label{gp}
\end{eqnarray}
Here $m$ stands for the particle mass, $\mu$ denotes the chemical potential, 
$V(\mathbf{r} -\mathbf{r}')$ represents an arbitrary two-body
potential, while $U(\mathbf{r})$ describes an external disorder potential. Denoting the disorder ensemble average as $\left\langle \,\bullet\,\right\rangle$, a homogeneous disordered system
can be described, without any loss of generality, by a vanishing 
mean value $\left\langle U(\mathbf{r})\right\rangle = 0$ and an arbitrary correlation function
\begin{equation}
\left\langle U(\mathbf{r})U(\mathbf{r}')\right\rangle = R(\mathbf r-\mathbf r')\, .
\end{equation}
In this section we present in detail a perturbative theory, which was developed earlier in Ref.~\cite{krumnow2011}, and calculate disorder corrections to the order parameter, 
the condensate depletion, the chemical potential, the superfluid depletion, and the sound velocity. The range of validity of this perturbation theory is limited by standard 
requirements for a mean-field approach: dilute, weakly-interacting BEC at low temperatures, when quantum fluctuations can be neglected. The perturbation expansion is performed 
with respect to the disorder strength, thus the disorder potential has to be sufficiently small compared to the chemical potential, i.e. $U(\mathbf{r})\ll \mu$.

We start with the observation that the GP Eq.~(\ref{gp}) represents a stochastic nonlinear partial differential
equation, where
the statistics of the condensate wave
function $\psi(\mathbf{r})$ is governed by the statistics of the disorder potential $U(\mathbf{r})$ \cite{navez2007}. Since $\psi(\mathbf r)$ describes the macroscopic occupation of the ground state, we assume it to be real without loss of generality.
In addition to the statistical properties of the random potential we will also assume that the macroscopic value of some physical quantity $A_{\textrm{mac}}$, obtained by coarse-graining of a microscopic quantity $A(\mathbf r)$ over a large volume $V$, gives the same result as the disorder ensemble average, namely:
\begin{eqnarray}\label{cg}
A_{\textrm{mac}}&=&\frac{1}{V}\int_{V}d^3 r\, A(\mathbf r)=\left\langle A\right\rangle\,.
\end{eqnarray} 
Here the length of the coarse-graining $\ell\sim V^{1/3}$ is assumed to be larger than both the correlation length $\sigma$ of the disorder potential $U(\mathbf{r})$, and the healing length $\xi=\hbar/\sqrt{2 m n g}$, which represents the characteristic distance at which the condensate wave-function responds to some perturbation in the external potential:
\begin{eqnarray}
 \ell&\gg&\sigma,\xi.
\end{eqnarray}
In the definition of the healing length $n$ represents the density of the fluid and $g = 4\pi \hbar^2 a_s/m$ denotes the strength of the short-range interaction part of the two-particle interaction potential $V(\mathbf{r}-\mathbf{r}')=g\delta(\mathbf{r}-\mathbf{r}')+\ldots$, expressed in terms of the s-wave scattering length $a_s$.

We consider the case of a sufficiently small random potential $U(\mathbf{r})\ll \mu\approx gn$, when the perturbative decomposition of the wave function of the system is justified:
\begin{eqnarray}
\label{expansion}
\psi(\mathbf{r}) = \psi_0 (\mathbf{r}) + \psi_1(\mathbf{r}) + \psi_2(\mathbf{r})+\ldots\,,
\end{eqnarray}
where $\psi_l(\mathbf{r})$ corresponds to the correction of the wave function of order $l$ in the disorder. 
Solving the GP equation (\ref{gp}) in the zeroth order of $U(\mathbf{r})$ gives
\begin{equation}\label{eq:pert0}
\psi_0^2=\frac{\mu}{V(\mathbf{k}=0)}\,,
\end{equation}
whereas the first order correction is straight-forwardly calculated and its Fourier transform reads
\begin{equation}\label{eq:pert1}
\psi_1(\mathbf{k})=-\frac{\psi_0 U(\mathbf{k})}{\frac{\hbar^2 k^2}{2m}+2\psi_0^2V(\mathbf{k})}\,.
\end{equation}
We note that its disorder ensemble average vanishes. Therefore, we also have to determine the second-order result, which turns out to be
\begin{widetext}
\begin{equation}\label{eq:pert2}
\psi_2(\mathbf{k})=-\int\frac{d^3 k'}{(2\pi)^3}\, 
\frac{U(\mathbf{k}-\mathbf{k}')\psi_1(\mathbf{k}')+
\psi_0 [2V(\mathbf{k}')+V(\mathbf{k})]\psi_1(\mathbf{k}')\psi_1(\mathbf{k}-\mathbf{k}')}
{\frac{\hbar^2 k^2}{2m}+2\psi_0^2V(\mathbf{k})}\, .
\end{equation}
\end{widetext}
The results obtained in Refs.~\cite{sanchez-palencia2006,gaul2011} can be considered as special cases of the above general approach. For instance,
we point out that Sec.~II of Ref.~\cite{gaul2011} contains a discrete version of Eqs.~(\ref{expansion})--(\ref{eq:pert1}) for the case of pure contact interaction, which is here generalized to an arbitrary two-body interaction. Note that the second-order correction (\ref{eq:pert2}) is slightly different since we take the chemical potential to be constant, whereas in Ref.~\cite{gaul2011} the density is taken to be constant.

In the following subsections we use the above outlined systematic perturbative approach \cite{krumnow2011} and calculate several physical properties of the dirty BEC and 
their respective disorder corrections.

\subsection{Order parameter and condensate depletion}

In analogy to quantum field theory, the one-particle density matrix is defined as $\left\langle \psi(\mathbf{r})\psi(\mathbf{r}')\right\rangle$ \cite{graham2009}.
The macroscopic fluid density is the diagonal part of the one-particle density matrix according to $n=\left\langle \psi^2(\mathbf{r})\right\rangle$,
whereas the condensate density is usually defined as the off-diagonal long-range order (ODLRO) parameter via \cite{graham2009}
\begin{equation}\label{eq:n0def}
n_0=\lim_{|\mathbf{r}-\mathbf{r^\prime}|\rightarrow \infty}\left\langle \psi(\mathbf{r})\psi(\mathbf{r}')\right\rangle\,.
\end{equation}
Performing the coarse-graining of the one-particle density matrix $\left\langle \psi(\mathbf{r})\psi(\mathbf{r}')\right\rangle$ over the fixed volume $V$ before taking the limit does not change the result
\begin{equation}
n_0=\lim_{|\mathbf{r}-\mathbf{r^\prime}|\rightarrow \infty} \frac{1}{V^2}\int_{V\otimes V} d ^3r_1 d ^3r_2\, \left\langle \psi(\mathbf{r}+\mathbf{r}_1)\psi(\mathbf{r}'+\mathbf{r}_2)\right\rangle\,.
\end{equation}
The integration commutes with the disorder ensemble  average and using Eq.~(\ref{cg}), we obtain
\begin{equation}\label{n0}
n_0=\left\langle \left\langle \psi(\mathbf{r})\right\rangle\left\langle \psi(\mathbf{r}')\right\rangle\right\rangle=\left\langle \psi(\mathbf{r})\right\rangle^2\,.
\end{equation}
The last equality follows from the fact that the average of an already averaged expression can be omitted. Therefore, the depletion of the condensate due to disorder, 
which is defined as $n-n_0=\left\langle \psi^2\right\rangle-\left\langle \psi\right\rangle^2$, is simply identified with the variance of the wavefunction.
Physically, this condensate depletion is due to the formation of fragmented
condensates in the respective local minima
of the random potential. Defining a separate Bose-glass order parameter by considering the ODLRO parameter of the two-particle density matrix \cite{graham2009}
\begin{eqnarray}\label{q}
(n_0+q)^2 =
\lim_{|\mathbf{r}-\mathbf{r^\prime}|\rightarrow \infty} \left\langle \psi(\mathbf{r})^2 \psi
(\mathbf{r^\prime}) ^2\right\rangle=n^2\,,
\end{eqnarray}
shows that the density of the fragmented
condensates~$q$ defined in Eq.~(\ref{q}) coincides with
the condensate depletion $n - n_0$. To this end the disorder ensemble average is obtained along the same lines as Eqs.~(\ref{eq:n0def})--(\ref{n0}). Thus, we
conclude that the localization phenomenon for
quenched disorder follows already from a mean-field description of the dirty boson
problem. 
Therefore, our mean-field approach 
represents a simplified derivation
of the disorder-induced condensate depletion
in comparison with the Bogoliubov theory of
Refs.~\cite{huang1992,giorgini1994,kobayashi2002,lugan2007,falco2007,hu2009,gaul2011,lugan2011,muller2012}.
Note that disorder effects on Bogoliubov quasiparticles
have recently been analyzed in Refs.~\cite{gaul2011,muller2012}.

The perturbative expansion (\ref{expansion}) now yields for the particle density
\begin{equation}
n = \left\langle \psi(\mathbf{r})^2\right\rangle=\psi_0^2+\left\langle \psi_1(\mathbf{r})^2\right\rangle+2\psi_0\left\langle \psi_2(\mathbf{r})\right\rangle+\ldots\,,\label{eq:dens}
\end{equation}
and, correspondingly, for the condensate density
\begin{equation}
n_0 = \left\langle \psi(\mathbf{r})\right\rangle^2=\psi_0^2+2\psi_0\left\langle \psi_2(\mathbf{r})\right\rangle+\ldots\,.
\end{equation}
With this the condensate depletion results to be
\begin{equation}
n-n_0 = \left\langle \psi_1(\mathbf{r})^2\right\rangle+\ldots \, .
\end{equation}
Using Eq.~(\ref{eq:pert1}) we arrive at the following expression:
\begin{eqnarray}\label{eq:deplgen}
n-n_0=  n \int \frac{d^3 k}{(2 \pi)^3}\frac{R(\mathbf{k})}{\left[\frac{\hbar^2 k^2}{2m}+2nV(\mathbf{k})\right]^2}+\ldots\,.
\end{eqnarray}
Note that this represents a result for the condensate depletion in second order of the disorder potential for an arbitrary two-particle interaction potential and an arbitrary disorder correlation
function. Specializing to the delta-correlated disorder $R(\mathbf{k})=R$ and the contact interaction $V(\mathbf{k})=g$, Eq.~(\ref{eq:deplgen}) reduces to
\begin{equation}\label{eq:hm}
n-n_0=n_{\mathrm{HM}}=\frac{m^\frac{3}{2}R\sqrt{n}}{4\pi\hbar^3\sqrt{g}}\,,
\end{equation}
which is the seminal result originally obtained by Huang and Meng  \cite{huang1992} within the Bogoliubov theory of dirty bosons.

\subsection{Equation of state}

Solving the equation $\left\langle \psi^2(\mu_b)\right\rangle=n(\mu_b)$ for the chemical potential $\mu_b$ yields its dependence on the average fluid density $\mu_b=\mu_b(n)$. We have introduced the notation $\mu_b$, denoting the "bare" chemical potential, because it diverges for uncorrelated disorder regardless of the density $n$, as can be seen from inserting expressions (\ref{eq:pert0})--(\ref{eq:pert2}) into the second-order correction (\ref{eq:dens}):
\begin{equation}
\mu_b=nV(\mathbf{k}=0)-
\int\frac{d^3 k}{(2 \pi)^3}\frac
{\frac{\hbar^2k^2}{2m}R(\mathbf{k})}
{\left[\frac{\hbar^2k^2}{2m}+2nV(\mathbf{k})\right]^2}+\ldots\,.
\end{equation}
This unphysical ultraviolet divergence can be removed by renormalizing the chemical potential \cite{falco2007}. If the density of the system vanishes, i.e. if there are no particles in the system, the energy needed for a particle to be added also has to vanish $\mu(n=0)=0$. Therefore, we define the renormalized chemical potential according to
\begin{equation}\label{ren}
\mu(n)=\mu_b(n)-\mu_b(0).
\end{equation}
With this we obtain in second order of the disorder strength the renormalized chemical potential:
\begin{eqnarray}
&&\hspace*{-10mm}
\mu=nV(\mathbf{k}=0)\label{eq:eqstgen}\\
&&\hspace*{-3mm} + 4n\int\frac{d^3 k}{(2 \pi)^3}\frac
{V(\mathbf{k})R(\mathbf{k})\left(\frac{\hbar^2k^2}{2m}+nV(\mathbf{k})\right)}
{\frac{\hbar^2k^2}{2m}\left[\frac{\hbar^2k^2}{2m}+2nV(\mathbf{k})\right]^2}+\ldots\,,
\nonumber
\end{eqnarray}
which does not contain an ultraviolet divergence.

For calculating the sound velocity later on we will also need the expression for the compressibility of the fluid, or its inverse given by $\partial \mu/\partial n$. Note that from Eq.~(\ref{ren}) it follows that the obtained result does not depend on whether we use $\mu$ or $\mu_b$. Thus, from the perturbative expansion (\ref{eq:eqstgen}) we read off:
\begin{equation}\label{eq:dmdngen}
\frac{\partial\mu}{\partial n} =
V(\mathbf{k}=0)+4\int \frac{d^3 k}{(2 \pi)^3}
\frac{\frac{\hbar^2 k^2}{2m}R(\mathbf{k})V(\mathbf{k})}{\left[\frac{\hbar^2 k^2}{2m}+2nV(\mathbf{k})\right]^3}+\ldots\,.
\end{equation}

\subsection{Superfluidity}
\label{sec:sd}

Without disorder and at $T=0$, the whole system is in a superfluid state, moving with an arbitrary wavevector $\mathbf{k}_S$, which corresponds to the superfluid velocity $\mathbf{v}_S=\hbar\mathbf{k}_S/m$. By introducing disorder that moves with the velocity $\hbar\mathbf{k}_U/m$, some part of the fluid will be moving together with it. The normal, i.e. non-superfluid, component of the fluid $n_N$ is defined as the part that moves together with the disorder, while the superfluid component $n_S$ is defined as the fraction of the fluid that moves with the superfluid wavevector $\mathbf{k}_S$. Therefore, the macroscopic current density $\left\langle \mathbf{j}(\mathbf{r})\right\rangle$ can be separated in this two-fluid picture as follows:
\begin{equation}\label{curr}
\left\langle \mathbf{j}(\mathbf{r})\right\rangle=n_S \mathbf{k}_S+n_N\mathbf{k}_U\, .
\end{equation}
The averaged current density $\left\langle \mathbf{j}(\mathbf{r})\right\rangle$ can be obtained by analyzing the underlying time-dependent GP equation for the system:
\begin{widetext}
\begin{equation}
\label{supgp}
\left[
-\frac{\hbar^2}{2m}\Delta+
U\left(\mathbf{r}-\mathbf{k}_U \frac{\hbar}{m}t\right)+\int d^3r'\, V(\mathbf{r}-\mathbf{r'})\Psi_S^*(\mathbf{r'},t)\Psi_S(\mathbf{r'},t)
\right]\Psi_S(\mathbf{r},t)
=i\hbar\frac{\partial\Psi_S(\mathbf{r},t)}{\partial t}\, ,
\end{equation}
\end{widetext}
where the condensate wave function $\Psi_S$ is a product of some as yet unknown function $\psi_S$ and a plane wave with wavevector $\mathbf{k}_S$ that corresponds to the clean-case solution:
\begin{equation}\label{supsol}
\Psi_S(\mathbf{r},t)=e^{i \mathbf{k}_S \mathbf{r}}
\psi_S(\mathbf{r},t) e^{-\frac{i}\hbar\left(\mu+\frac{\hbar^2k_S^2}{2m}\right)t}.
\end{equation}
Substituting the ansatz (\ref{supsol}) into Eq.~(\ref{supgp}), changing variables via $\mathbf{x}=\mathbf{r}-\mathbf{k}_U \frac{\hbar}{m}t$ and introducing $\mathbf{K}=\mathbf{k}_S-\mathbf{k}_U$ leads to
\begin{widetext}
\begin{equation}\label{gpx}
\left[
-\frac{\hbar^2}{2m}\Delta-i\frac{\hbar^2}{m}\mathbf{K}\cdot\nabla+
U(\mathbf{x})-\mu+\int d^3 x'\,  V(\mathbf{x}-\mathbf{x'})\psi_S^*(\mathbf{x'})\psi_S(\mathbf{x'})
\right]
\psi_S(\mathbf{x})=0\, .
\end{equation}
\end{widetext}
Although $\psi_S$ should in general depend on $t$, it can be shown via mathematical induction on the perturbative solution that all orders of $\psi_S(\mathbf{x},t)$ turn out to be time-independent \cite{nikolic2012}. Note that $\psi_S$ does not depend explicitly on the wavevectors $\mathbf{k}_S$ and $\mathbf{k}_U$, but only on their difference $\mathbf{K}$. 
Here, we are only interested in small values of $\mathbf{K}$ and, therefore, perform the expansion $\psi_S=\psi+\mathbf{p}\mathbf{K}+\ldots$, with $\mathbf{p}=(\partial\psi_S/ \partial {\mathbf{K}})_{\mathbf{K}=\mathbf{0}}$. An explicit equation for $\mathbf{p}$ can be obtained by performing the derivative of Eq.~(\ref{gpx}) with respect to $\mathbf{K}$, yielding 
\begin{widetext}
\begin{equation}
-\frac{\hbar^2}{2m}\Delta\mathbf{p}(\mathbf{x})-\frac{i\hbar^2}{m}\nabla\psi(\mathbf{x})+\left[U(\mathbf{x})-\mu\right]\mathbf{p}(\mathbf{x})
+\int d^3 x'\, V(\mathbf{x}-\mathbf{x'})\Big\{[\mathbf{p}^*(\mathbf{x'})+\mathbf{p}(\mathbf{x'})]\psi(\mathbf{x'})\psi(\mathbf{x})+\psi(\mathbf{x'})^2\mathbf{p}(\mathbf{x})
\Big\}=0\,.
\label{eqp}
\end{equation}
\end{widetext}
If we take into account Eq.~(\ref{supsol}), the standard definition of the  current density,
\begin{equation}
\left\langle \mathbf{j}\right\rangle = \frac{1}{2i}\left\langle \Psi^*_S\nabla\Psi_S-\Psi_S\nabla\Psi^*_S\right\rangle\, ,
\end{equation}
transforms into
\begin{equation}
\left\langle \mathbf{j}\right\rangle = \left\langle \psi_S^*\psi_S\right\rangle\mathbf{k}_S+\frac{1}{2i}\left\langle \psi^*_S\nabla\psi_S-\psi_S\nabla\psi^*_S\right\rangle\, ,
\end{equation}
which then can be further reduced to
\begin{equation}
\left\langle \mathbf{j}\right\rangle  = n\mathbf{k}_S+\left(\left\langle \psi \nabla\otimes\mathrm{Im}\, \mathbf{p}\right\rangle-\left\langle \nabla\psi\otimes\mathrm{Im}\, \mathbf{p}\right\rangle\right)\mathbf{K}+\ldots\,.\label{jlexp}
\end{equation}
In the last line we have neglected higher than linear orders in $\mathbf{k}_U$ and $\mathbf{k}_S$. 

For small disorder strengths, we expand Eq.~(\ref{jlexp}) with respect to $U$ up to second order. To this end we take into account the homogeneity of our problem, that leads to $\partial_i\left\langle \mathbf{p}_2(\mathbf{x})\right\rangle=0$, and note that in zeroth order $\psi_S$ does not depend on $\mathbf{K}$, thus leading to $\mathbf{p}_0=0$. 
With this we obtain
\begin{equation}
\left\langle \mathbf{j}\right\rangle= n\mathbf{k}_S+(\left\langle \psi_1 \nabla\otimes\mathrm{Im}\, \mathbf{p}_1\right\rangle-\left\langle \nabla\psi_1\otimes\mathrm{Im}\, \mathbf{p}_1\right\rangle)\mathbf{K}+\ldots\,,\label{jlexpu}
\end{equation}
where also $\mathbf{p}$ is expanded in the disorder strength according to $\mathbf{p}=\mathbf{p}_0+\mathbf{p}_1+\mathbf{p}_2+\ldots\,$.
Solving the imaginary part of Eq.~(\ref{eqp}) in first order in $U$ yields the Fourier transform of $\mathrm{Im}\, \mathbf{p}_1$:
\begin{equation}\label{eq:fimp1}
(\textrm{Im}\,\mathbf{p}_{1})(\mathbf{k})=2i\frac{\mathbf{k}}{k^2}\psi_1(\mathbf{k})\,.
\end{equation}
Thus, together with the solution for $\psi_1$ given by Eq.~(\ref{eq:pert1})  and a comparison with Eq.~(\ref{curr}), 
we obtain from Eq.~(\ref{jlexpu}) the normal fluid density in the form
\begin{equation}
\hat{n}_N=4n \int \frac{d^3 k}{(2 \pi)^3}\frac{\mathbf{k}\otimes\mathbf{k}}{k^2}\frac{R(\mathbf{k})}{\left[\frac{\hbar^2 k^2}{2m}+2nV(\mathbf{k})\right]^2}+\ldots\,.\label{eq:sdgen}
\end{equation}
Note that in general the non-superfluid component is represented by a tensor \cite{ueda2010}.
 
In the case of a cylindrically symmetric system, we can choose the  symmetry axis as the $z$-axis and  denote the polar and the azimuth angle by $\theta$ and $\varphi$, so integrating Eq.~(\ref{eq:sdgen}) in spherical coordinates with respect to $\varphi$ yields the angle dependence
\begin{widetext}
\begin{eqnarray}
\sin\theta \int_{0}^{2\pi}d\varphi~ \mathbf{e}_\mathbf{k}\mathbf{e}_\mathbf{k}^T&=&\sin{\theta}\int_{0}^{2\pi}d\varphi
\left(\begin{array}{ccc}
\sin ^2\theta\cos ^2\varphi  & \sin ^2\theta\sin \varphi\cos \varphi & \sin \theta\cos \theta\cos \varphi\\
  \sin ^2\theta\sin \varphi\cos \varphi & \sin ^2\theta\sin ^2\varphi & \sin \theta\cos \theta\sin \varphi \\
\sin \theta\cos \theta\cos \varphi &  \sin \theta\cos \theta\sin \varphi  & \cos ^2\theta
\end{array} \right)\nonumber\\
\label{eq:cyl}
&&= \sin\theta\left(\begin{array}{ccc}
\pi(1-\cos ^2\theta)  & 0 & 0\\
0 & \pi(1-\cos ^2\theta) &0 \\
0 & 0 & 2\pi\cos ^2\theta
\end{array} \right)\,.
\end{eqnarray}
\end{widetext}
If both $V(\mathbf{k})$ and $R(\mathbf{k})$ are $\theta$-independent, i.e. if we have spherical symmetry, integrating Eq.~(\ref{eq:cyl})  with respect to $\theta$ leads to a solution in second order of the disorder potential \cite{huang1992,giorgini1994,kobayashi2002,lugan2007,falco2007,hu2009,gaul2011,lugan2011,muller2012}:
\begin{equation}\label{eq:isot}
\hat{n}_N=\frac{4}{3}(n-n_0)\hat I\,.
\end{equation}
This result shows that the superfluid depletion will be larger by a factor of $4/3$ than the condensate depletion. Thus, the localized fragmented part of the fluid hinders the superfluid to move.

\subsection{Sound velocity}

In the mean-field approach, we can also define the sound velocity by perturbing the time-independent solution with a small time-dependent variation. It is expected that sound waves with wavelengths of the order of the correlation length would scatter and interfere due to disorder hills and valleys, making the sound velocity impossible to define precisely. Locally, the sound waves would have the same speed as in the clean case. For sound waves with wavelengths much larger than the disorder correlation length, the sound velocity can be calculated using the hydrodynamical approach \cite{krumnow2011}.
Hydrodynamic equations are valid in the macroscopic regime and can only be used for slowly varying quantities that do not depend on the specific microscopic realization. Spatial averaging over distances much larger than the correlation length and much smaller than the wavelength solves the problem. Assuming that it gives the same result as the disorder ensemble average, we obtain the hydrodynamic equations for the macroscopic, i.e. disorder averaged, quantities in the form
\begin{eqnarray}
\frac{\partial n(\mathbf{x},t)}{\partial t}+\nabla \left[\hat{n}_S(\mathbf{x},t)\mathbf{v}_S(\mathbf{x},t)\right]&=&0\,,\\
\hspace*{-5mm}m\frac{\partial \mathbf{v}_S(\mathbf{x},t)}{\partial t}+\nabla \left[\frac{m\mathbf{v}_S(\mathbf{x},t)^2}{2}+\mu(n(\mathbf{x},t))\right]&=&0\,,
\end{eqnarray}
where  $n$ denotes the macroscopic density, and the disorder velocity $\mathbf{k}_U$ is taken to be zero.
If we write densities and the superfluid velocity as sums of homogeneous equilibrium values and small variations,
\begin{eqnarray}
n(\mathbf{x},t) &=& n+\delta n(\mathbf{x},t),\\
\hat{n}_S(\mathbf{x},t) &=& \hat{n}_S+\delta\hat{n}_S(\mathbf{x},t),\\
\mathbf{v}_S(\mathbf{x},t) &=& \delta\mathbf{v}_S(\mathbf{x},t),
\end{eqnarray}
as well as neglect second-order terms in the variations, we get the following linearized system of equations:
\begin{eqnarray}
\frac{\partial \delta n(\mathbf{x},t)}{\partial t}+\nabla \left[\hat{n}_S\delta\mathbf{v}_S(\mathbf{x},t)\right]&=&0\,,\label{eq:hid1}\\
\frac{\partial \delta\mathbf{v}_S(\mathbf{x},t)}{\partial t}=-\frac{1}{m}\nabla\mu \left[n+\delta n(\mathbf{x},t)\right]&=&\nonumber\\
&&\hspace*{-40mm}-\frac{1}{m}\frac{\partial\mu}{\partial n}\nabla\delta n(\mathbf{x},t)\, .\label{eq:hid2}
\end{eqnarray}
Taking the time derivative of Eq.~(\ref{eq:hid1}) and substituting the expression for the superfluid velocity variation from Eq.~(\ref{eq:hid2}) we obtain the generalized wave equation
\begin{equation}
\frac{\partial^2 \delta n(\mathbf{x},t)}{\partial t^2}-\frac{1}{m}\frac{\partial \mu}{\partial n}\nabla\left[\hat{n}_S\nabla\delta n(\mathbf{x},t)\right]=0.
\end{equation}
From the above equation we deduce that the sound velocity in the direction of some unit vector $\mathbf{q}$  is given by 
\begin{equation}\label{eq:cgen}
c_{\mathbf{q}}^2=\frac{1}{m}\frac{\partial\mu}{\partial n}\mathbf{q}^T\hat{n}_S\mathbf{q}\,,
\end{equation}
where the tensorial property of the superfluid density has been taken into account. In order to further evaluate the sound velocity (\ref{eq:cgen}) for small disorder, the perturbative results for both the inverse compressibility (\ref{eq:dmdngen}) and the superfluid density following from (\ref{eq:sdgen}) have to be taken into account.

\section{Dipolar interaction and Lorentz-correlated disorder}
\label{sec:diplor}

In this section we will specialize the previously developed perturbative formalism and consider BEC systems in the presence of two different anisotropies, namely an anisotropic dipolar interaction between the analyzed particles and an anisotropic disorder potential. The latter is widely studied and physically motivated, for instance, by the anisotropy of the laser-speckle potential \cite{dainty1984,billy2008}. In order to obtain analytical results, we model the disorder correlation function by a cylindrically-symmetric Lorentzian in Fourier space
\begin{equation}\label{eq:corf}
R(\mathbf{k})=\frac{R}{1+\sigma_\rho^2 k_\rho^2+\sigma_z^2 k_z^2}\,.
\end{equation}
The lengths $\sigma_\rho$ and $\sigma_z$ denote the perpendicular and the parallel correlation length, respectively, and their experimentally realistic values are typically in a broad range from a few to several hundreds healing lengths $\xi$. The function (\ref{eq:corf}) is not physically realistic, but the corresponding results qualitatively coincide with the case of an isotropic Gaussian-correlated disorder, which was numerically calculated in Ref.~\cite{krumnow2011}. Therefore, we expect that all phenomena, that appear here, would also appear qualitatively for a true laser-speckle correlation function in a setup where it decays monotonously with distance. 

Assuming that the van der Waals
forces between the atoms can be approximated at low energies by an effective contact interaction, the
interaction potential in the presence of an external field, that aligns the dipoles in a direction $\mathbf{m}$, takes the form \cite{yi2000}
\begin{equation}\label{eq:veff}
V(\mathbf{r})=g\delta(\mathbf{r})+\frac{C_{\rm dd}}{4\pi r^3}\left[1-3\cos^2\phi (\mathbf{m},\mathbf{r})\right]\,,
\end{equation}
where $\phi (\mathbf{m},\mathbf{r})$ represents the angle between vectors $\mathbf{m}$ and $\mathbf{r}$,
and $C_{\rm dd}$ denotes the dipole-dipole interaction strength. In the case of magnetic dipoles $C_{\rm dd}=\mu_0 m^2$, with $\mu_0$ being the magnetic permeability and the magnetic dipole moment $m$, whereas for electric dipoles we have $C_{\rm dd}= d^2/\varepsilon_0$, with the vacuum permeability $\varepsilon_0$ and the electric dipole moment $d$.
Introducing the ratio of the dipole-dipole and the contact interaction
$
\epsilon=C_{\rm dd}/3g\,,
$
and taking the Fourier transform of the potential, we obtain \cite{goral2000}
\begin{equation}\label{eq:poten}
V(\mathbf{k})=g\left\{1+\epsilon \left[3 \cos^2\phi (\mathbf{m},\mathbf{k})-1\right]\right\}\,.
\end{equation}
The interaction ratio $\epsilon$ takes values as small as $0.008$ for $^{87}$Rb, while for $^{52}$Cr it is around $0.16$. For a BEC of heteronuclear molecules, its value would be much higher, namely of the order of $100$.

The Huang and Meng result \cite{huang1992} for the condensate depletion (\ref{eq:hm}) is linear in $R$, and therefore we will compare the relative change of physical quantities due to disorder to the relative change of the condensate density. To this end we will define a dimensionless disorder correction for each relevant quantity: condensate density, chemical potential, superfluid density, and sound velocity.
Corrections defined in this way are expressed in terms of only three parameters: the relative dipole-dipole interaction strength $\epsilon$, and the correlation lengths in units of the healing lengths, i.e. $z_{\rho,z}=\sqrt{2}\sigma_{\rho,z}/\xi$. We consider systems with an overall cylindrical symmetry, where the disorder symmetry axis is parallel to the direction of the dipoles. Otherwise the angle between them would be a fourth parameter that would have to be taken into account.

\begin{figure*}[!t]
  \includegraphics[width=6cm]{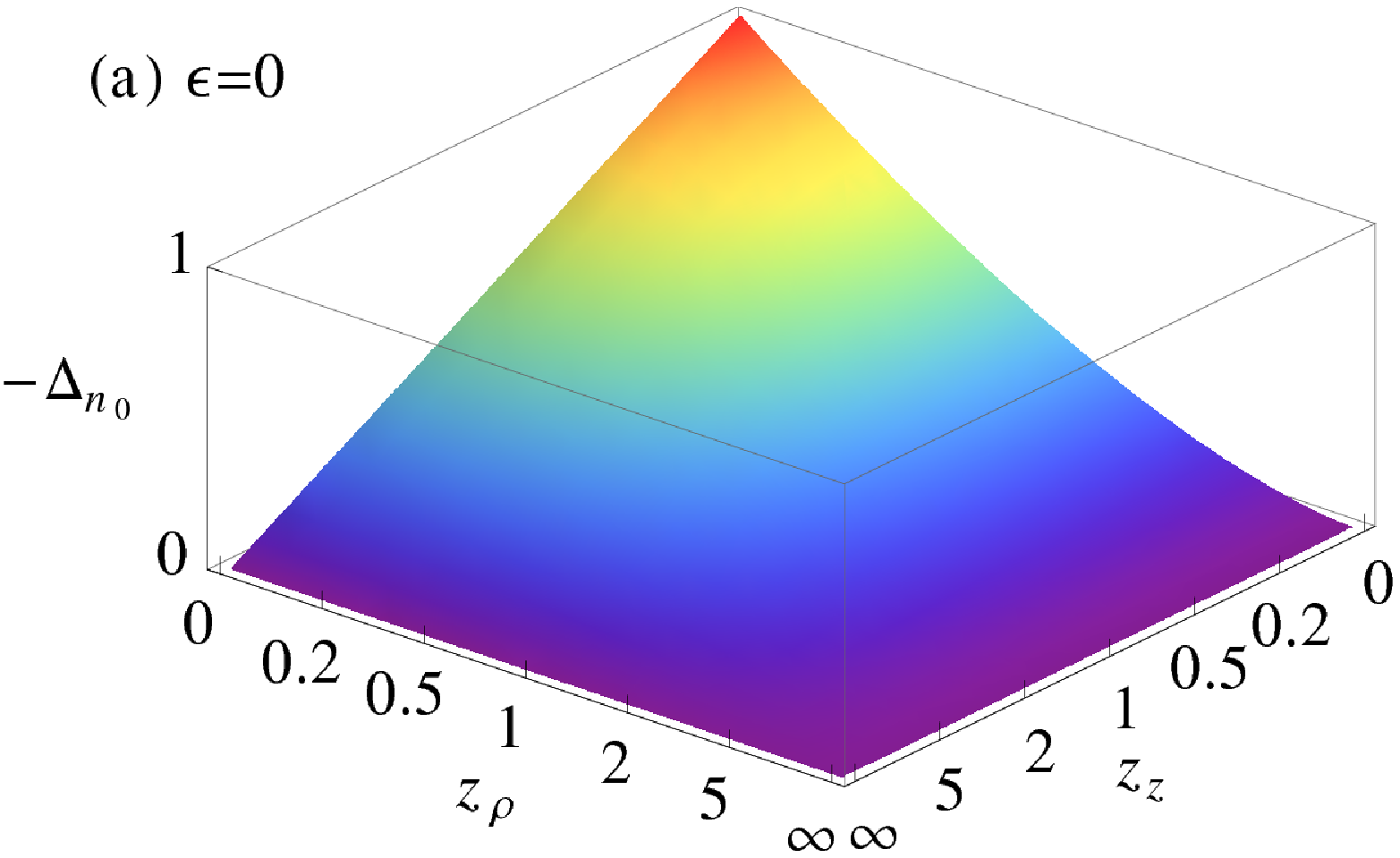}\hspace*{5mm}
  \includegraphics[width=6cm]{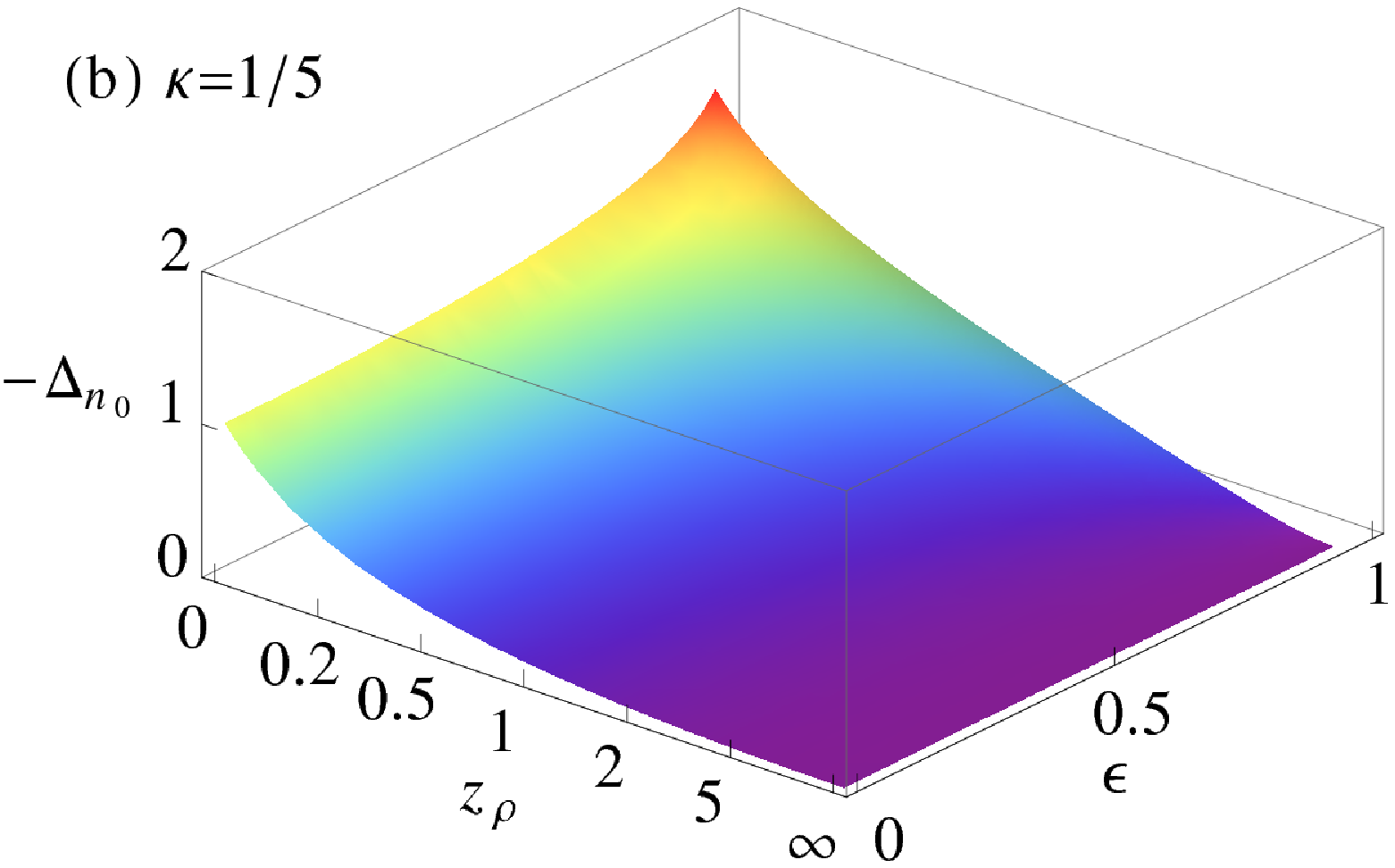}\vspace*{5mm}
  \includegraphics[width=6cm]{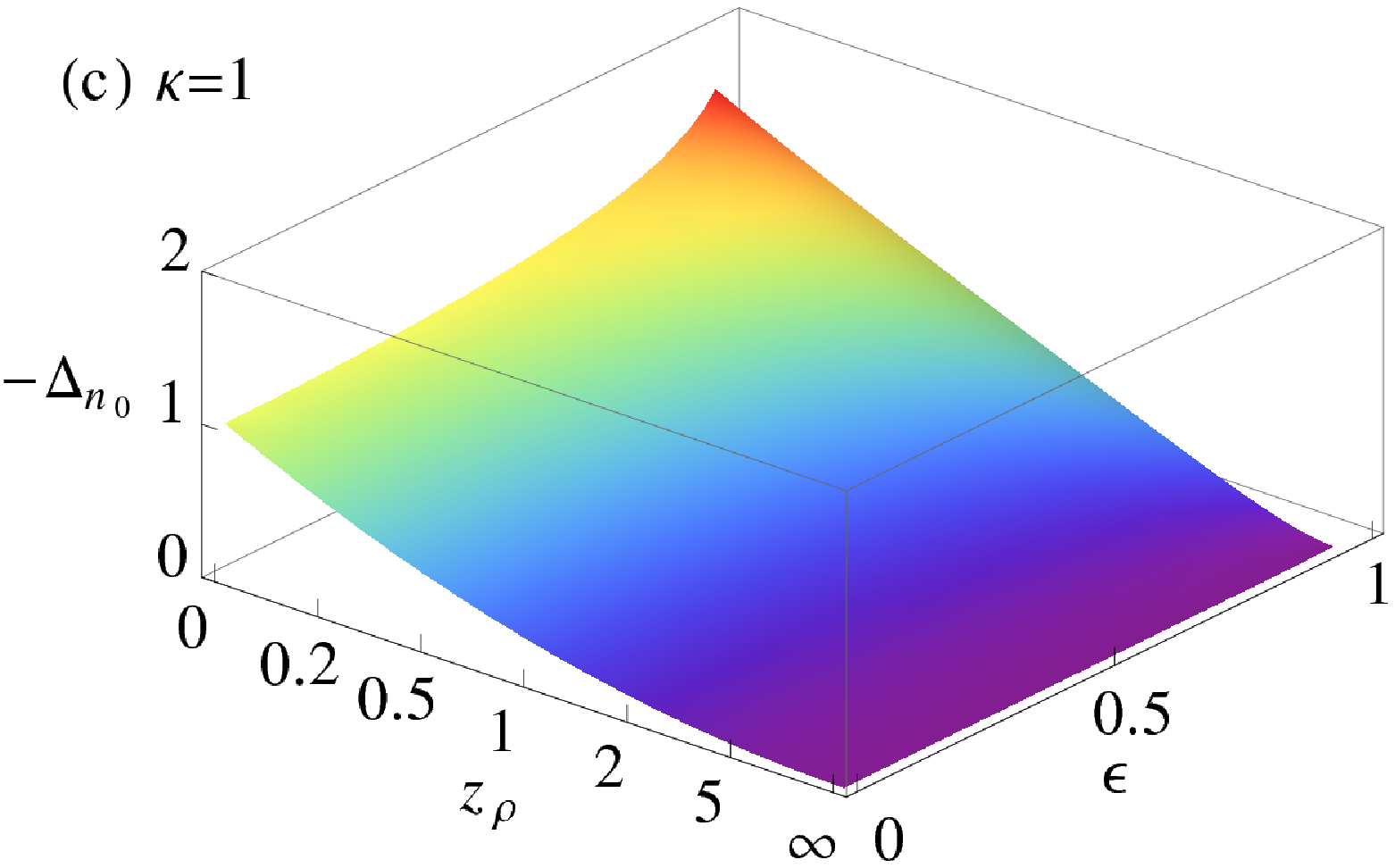}\hspace*{5mm}
  \includegraphics[width=6cm]{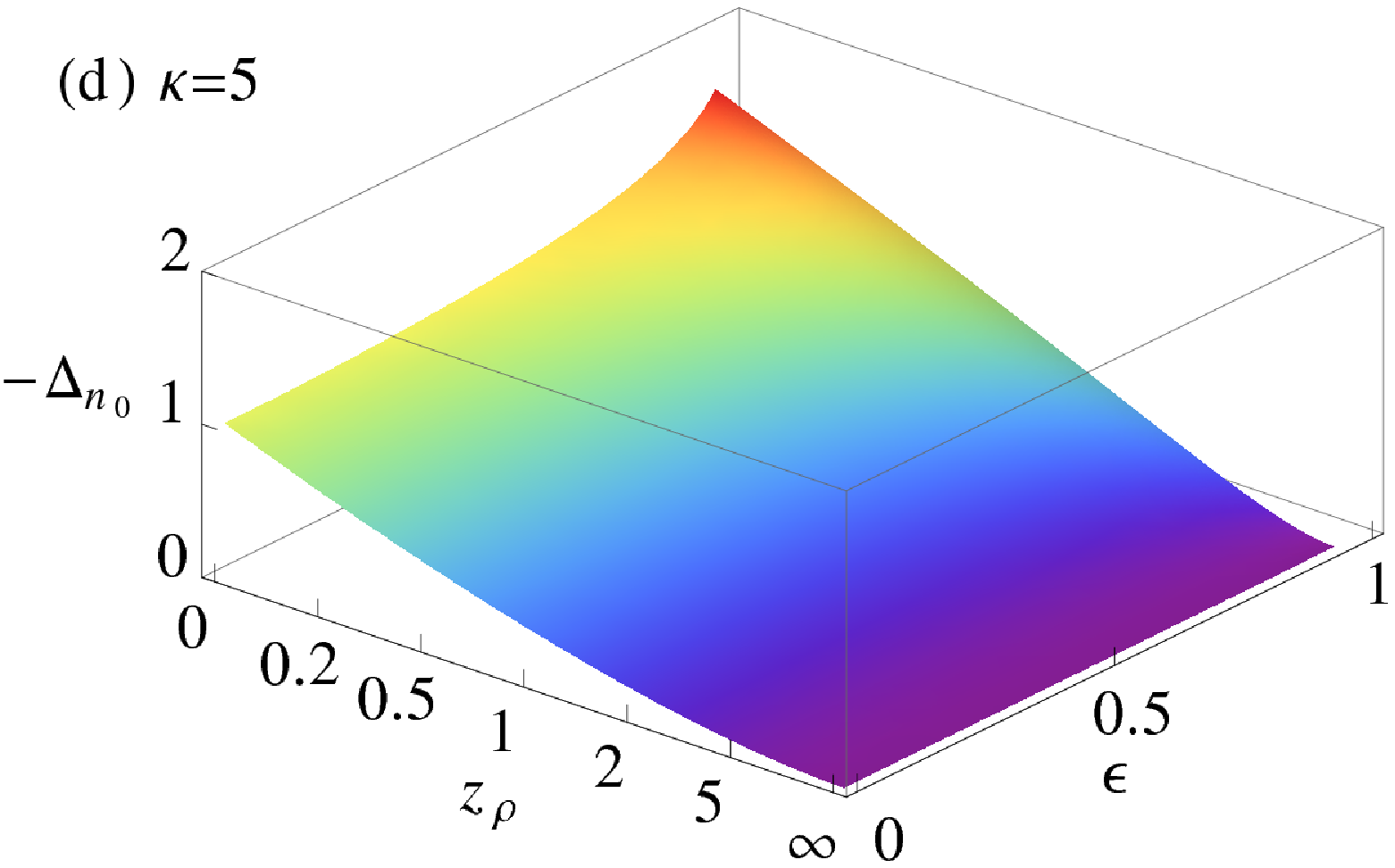}
  \caption{(Color online) Condensate depletion due to weak disorder: (a) as a function of the correlation lengths $z_\rho$ and $z_z$ for anisotropic disorder and pure contact interaction ($\epsilon=0$), expressed by Eq.~(\ref{eq:depletion});
(b) as a function of the correlation length $z_\rho$ and the ratio of the dipole-dipole and the contact interaction $\epsilon$ for anisotropic disorder with $\kappa=z_\rho/z_z=1/5$ (general expressions given in the Appendix); (c) $\kappa=1$, expressed by Eq.~(\ref{eq:depletion2}); (d) $\kappa=5$ (general expressions given in the Appendix).}
  \label{fig:n0}
\end{figure*}

The general case can be calculated analytically for all observables of interest, but the results are too cumbersome to be displayed here \cite{nikolic2012}, and we present them in the Appendix as well as in the Supplemental Material \cite{supplemental}. In this section we derive analytical results for the two special cases, namely the pure contact interaction with anisotropic disorder, and the dipolar interaction with isotropic disorder. We also present and discuss numerical results for the special cases, as well as for the general case with contact and dipole-dipole interaction
as well as anisotropic disorder.

%
%
%%%%%%%%%%%%%%%%%%%%%%%%%%%%%%%%%%%%%%%%%%%%%%%%%%%%%%%%%%%%%%%%%%
%%%%%%%%%%%%%%%%%%%%%%%%%%%%%%%%%%%%%%%%%%%%%%%%%%%%%%%%%%%%%%%%%%
%%%
%%%   condensate depletion
%%%
%%%%%%%%%%%%%%%%%%%%%%%%%%%%%%%%%%%%%%%%%%%%%%%%%%%%%%%%%%%%%%%%%%
%%%%%%%%%%%%%%%%%%%%%%%%%%%%%%%%%%%%%%%%%%%%%%%%%%%%%%%%%%%%%%%%%%
%

\subsection{Condensate depletion}

We now use the disorder correlation function and the interaction potential defined by Eqs.~(\ref{eq:corf}) and (\ref{eq:poten}) in order to calculate the disorder correction of the condensate density. To this end we introduce the dimensionless condensate depletion as follows:
\begin{equation}
\label{eq:deltan0def}
\Delta_{n_0}=\lim_{R\rightarrow0} \frac{n_0-n}{n_{\mathrm{HM}}}\, .
\end{equation}

Taking into account Eq.~(\ref{eq:hm}), making a substitution $\mathbf{k}\rightarrow \mathbf{k} \xi/\sqrt{2}$, denoting  the direction of the cylindrical symmetry of the disorder by $\mathbf{d}$ and introducing direction-dependent anisotropy functions $r$ and $v$ by
\begin{eqnarray}
r&=&\sqrt{z_\rho^2 \sin^2\phi (\mathbf{d},\mathbf{k})+z_z^2 \cos^2\phi (\mathbf{d},\mathbf{k})}\,,\label{eq:r}\\
v&=&\sqrt{1+\epsilon\left[3 \cos^2\phi (\mathbf{m},\mathbf{k})-1\right]}\,,\label{eq:v}
\end{eqnarray}
Eq.~(\ref{eq:deplgen}) yields the dimensionless value of the condensate depletion in second order in the form
\begin{equation}\label{eq:deplcor}
-\Delta_{n_0}=8\pi\int \frac{d^3 k}{(2 \pi)^3}\frac{1}{(\mathbf{k}^2+v^2)^2(1+\mathbf{k}^2 r^2)}\,.
\end{equation}
Assuming that the direction of dipoles is parallel to the disorder symmetry, i.e. $\mathbf{d}||\mathbf{m}$, the whole system will also become cylindrically symmetric. Writing Eq.~(\ref{eq:deplcor}) in spherical coordinates $(k,\theta,\varphi)$, integrating with respect to $k$ and $\varphi$ and changing the variable $t=\cos \theta$, leads to:
\begin{eqnarray}
-\Delta_{n_0}=\int_0^1 dt\frac{1}{v\left(1+v r\right)^2}\label{eq:deplcyl}
\end{eqnarray}
with functions $r$ and $v$ from Eqs.~(\ref{eq:r}) and (\ref{eq:v}) having the new form
\begin{eqnarray}
r&=&\sqrt{z_\rho^2+(z_z^2 -z_\rho^2)t^2}\,,\\
v&=&\sqrt{1-\epsilon+3\epsilon t^2}\,.
\end{eqnarray}
The two special cases, with pure contact interaction $(\epsilon=0, v=1)$ and with isotropic disorder $(z_\rho=z_z=r)$ can be solved explicitly using Euler substitutions $r=xt+z_\rho$ and $v=xt+\sqrt{1-\epsilon}$, respectively, in Eq.~(\ref{eq:deplcyl}), which leads to an integral of a rational function with respect to $x$. The analytic results for the two special cases are:
\begin{eqnarray}\label{eq:depletion}
&&\hspace*{-5mm}-\Delta_{n_0}\big|_{\epsilon=0}=
\frac{1}{(z_{\rho}-1)(z_{\rho}+1)}
\left[
\frac{2z_{\rho}^2}{(z_{\rho}+1)(z_{\rho}+z_z)}\right. \nonumber\\
&& \hspace*{5mm}\left. \times T\left(\frac{z_\rho-1}{z_\rho+1}\, \frac{z_z-z_\rho}{z_z+z_\rho}\right)
-\frac{1}{z_z+1}
\right]\,,\\
&&\hspace*{-5mm}-\Delta_{n_0}\big|_{z_\rho =z_z=z}=
\frac{z (1-\lambda )}{\left(-1+z^2 \delta ^2\right) [1-\lambda +z \delta  (1+\lambda )]}\nonumber\\
&&\hspace*{5mm}+\frac{(-1+\lambda )}{\delta  (-1+z
\delta ) (1+z \delta )^2}\, T\left(\frac{z \delta-1}{z\delta+1} \lambda \right)\, , \label{eq:depletion2}
\end{eqnarray}
where, for brevity, we introduced $\delta$ and $\lambda$ by $\delta^2=1-\epsilon$ and 
$\epsilon=\frac{4\lambda}{1-2\lambda+3\lambda^2}$, and $T(x)=\frac{\arctan\sqrt{x}}{\sqrt{x}}$ is a new function, well defined for positive values of $x$ and analytically continuable for $-1<x\leq 0$.

In Fig.~\ref{fig:n0} we have displayed the results for the condensate depletion. Figure~\ref{fig:n0}(a) corresponds to the special case of pure contact interaction ($\epsilon=0$), described by Eq.~(\ref{eq:depletion}), and Fig.~\ref{fig:n0}(c) corresponds to the special case of isotropic disorder ($\kappa=z_\rho/z_z=1$), given by Eq.~(\ref{eq:depletion2}). Note that for $\epsilon=0$ and $z_\rho=z_z=0$ the dimensionless condensate depletion is $1$, which coincides with the Huang and Meng result. For increasing correlation lengths the depletion decreases, and vanishes for infinite correlation length, as expected, since then the disorder is flat. The tiny asymmetry in Fig.~\ref{fig:n0}(a) comes from the fact that $z_\rho$ describes two spatial dimensions and, therefore, has a more pronounced effect on the depletion than $z_z$, which represents only one spatial dimension. As we see in Fig.~\ref{fig:n0}(c), increasing the relative dipole-dipole interaction leads to a larger depletion and, eventually, when it reaches the same order as the contact interaction, the BEC collapses, which corresponds to the divergence of the depletion for $\epsilon\to 1$.

\begin{figure*}[!t]
  \includegraphics[width=6cm]{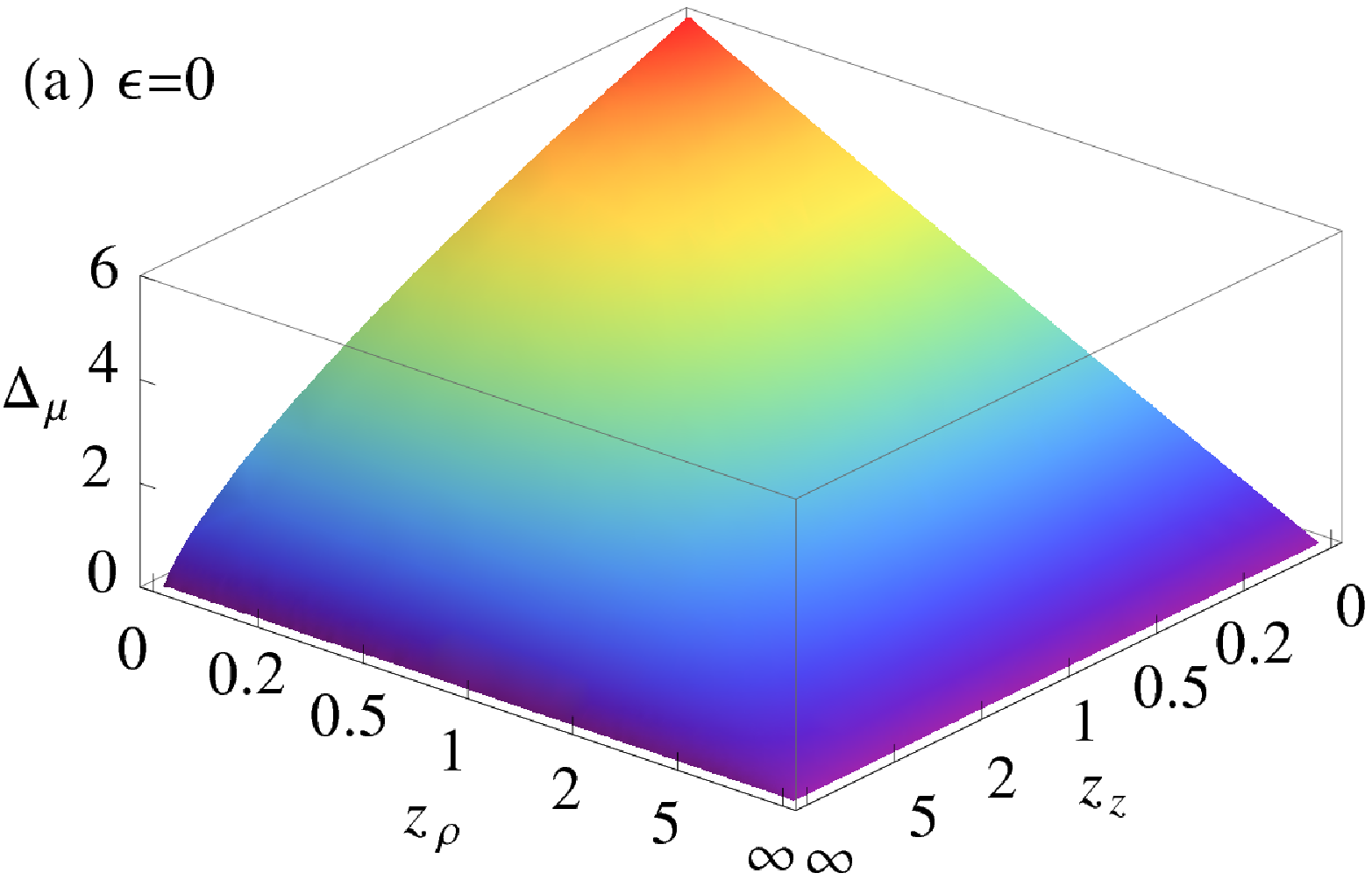}\hspace*{5mm}
  \includegraphics[width=6cm]{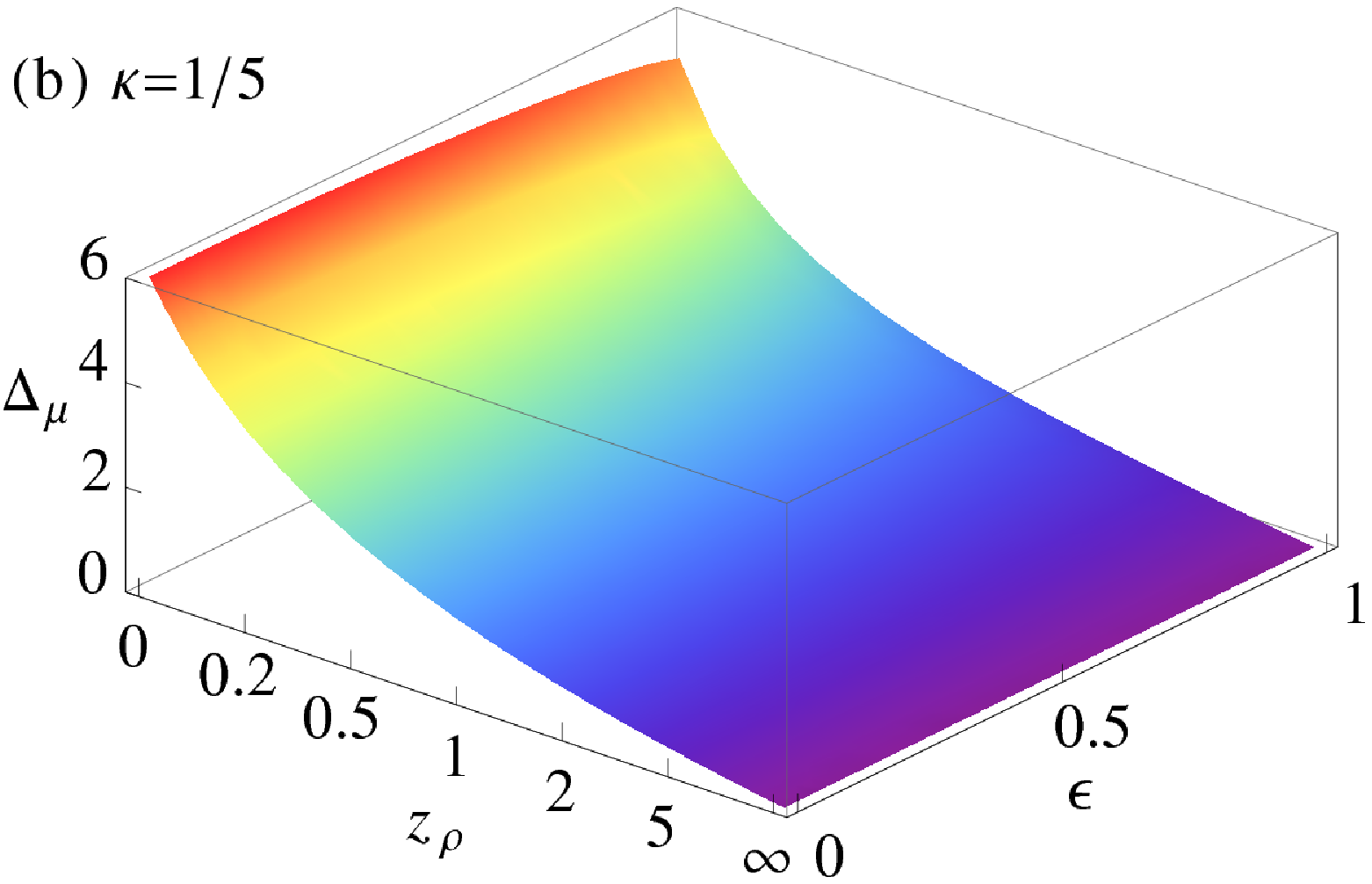}\vspace*{5mm}
  \includegraphics[width=6cm]{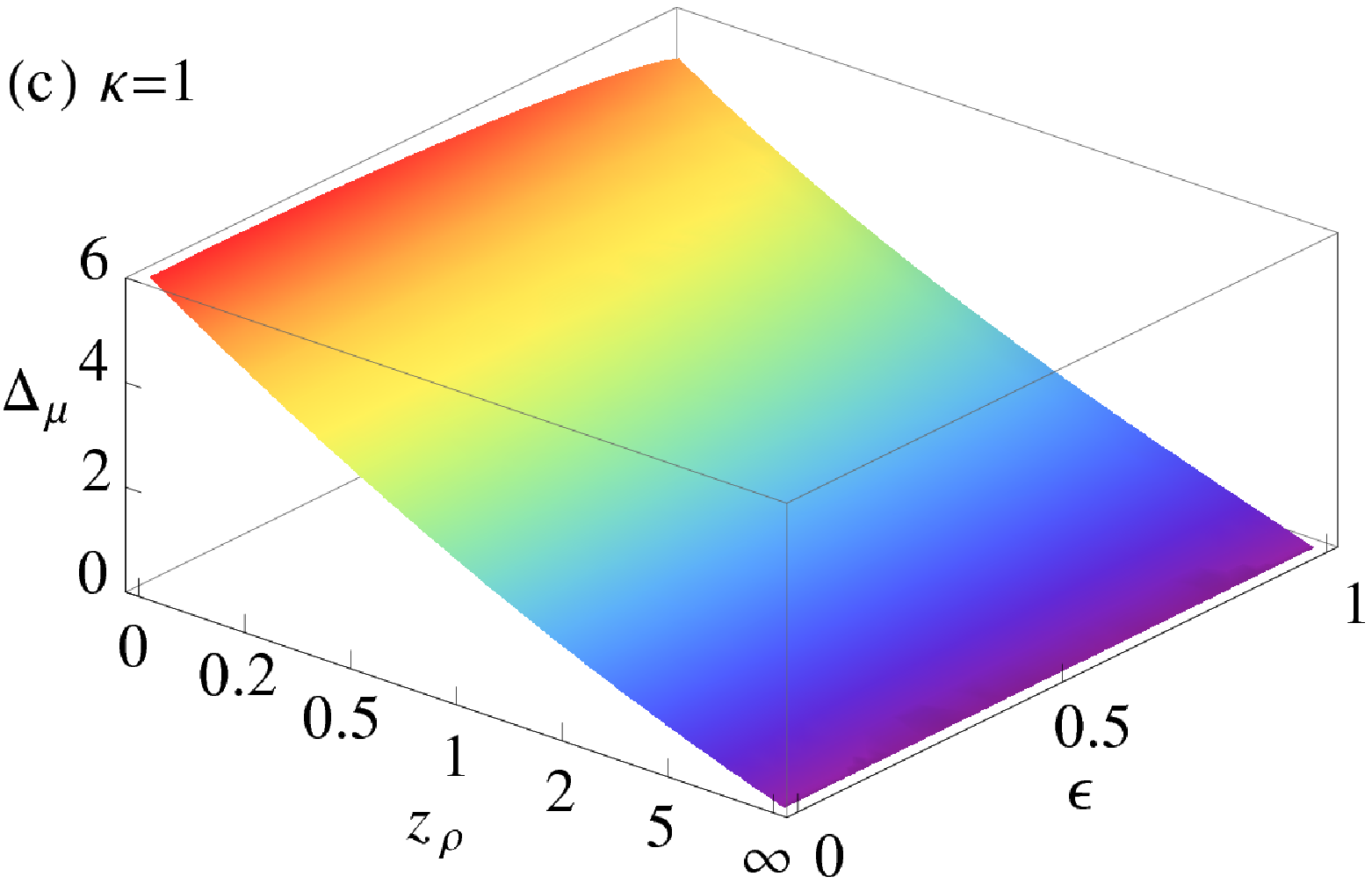}\hspace*{5mm}
  \includegraphics[width=6cm]{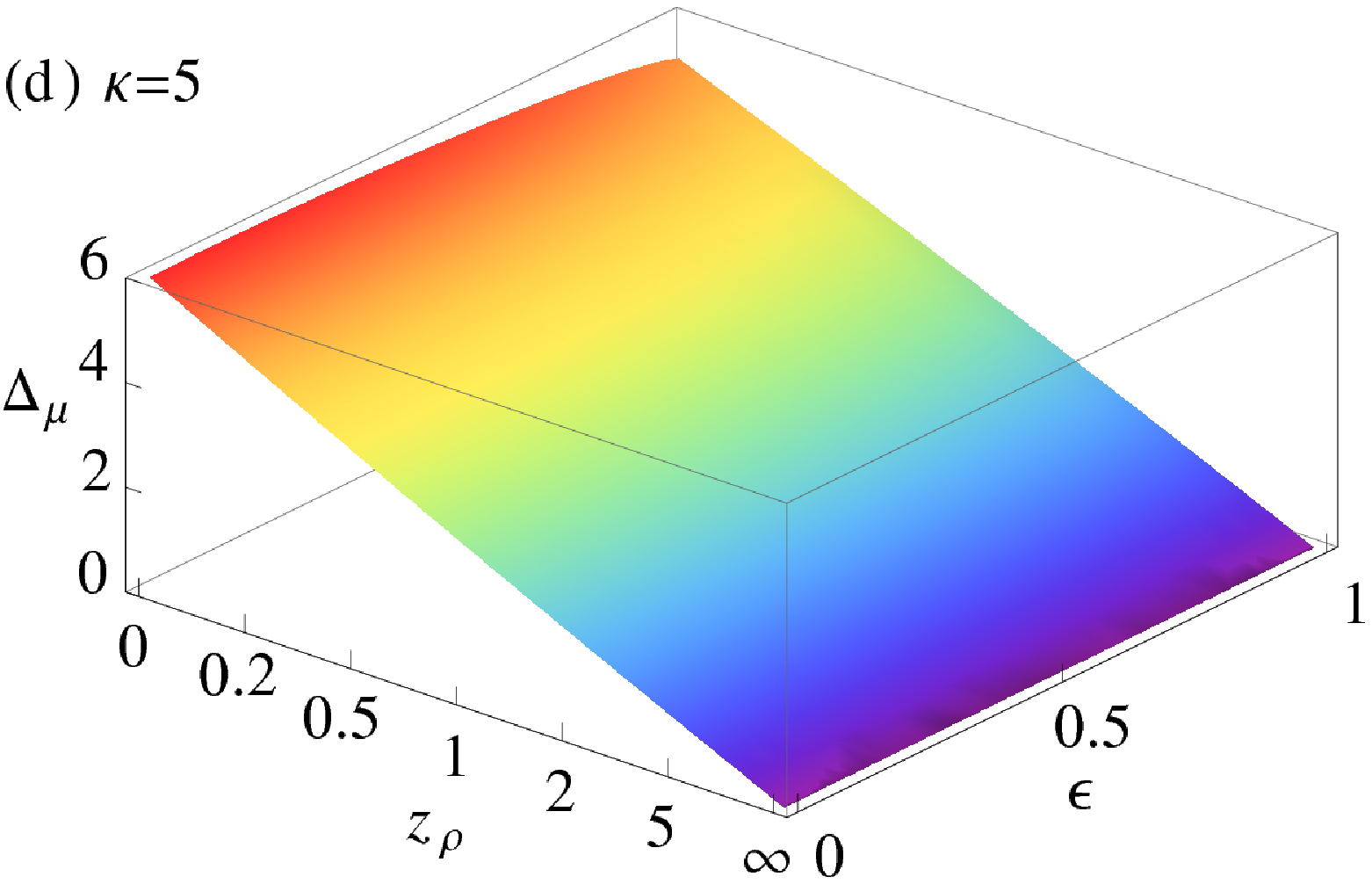}
  \caption{(Color online) Correction to the chemical potential due to weak disorder: (a) as a function of the correlation lengths $z_\rho$ and $z_z$ for anisotropic disorder and pure contact interaction ($\epsilon=0$), expressed by Eq.~(\ref{eq:eqstsol});
(b) as a function of the correlation length $z_\rho$ and the ratio of the dipole-dipole and the contact interaction $\epsilon$ for anisotropic disorder with $\kappa=z_\rho/z_z=1/5$ (general expressions given in the Appendix); (c) $\kappa=1$, expressed by Eq.~(\ref{eq:eqstsol2}); (d) $\kappa=5$ (general expressions given in the Appendix).}
    \label{fig:mu}
\end{figure*}

The general case with both contact and dipole-dipole interaction as well as anisotropic disorder is shown in Figs.~\ref{fig:n0}(b) and \ref{fig:n0}(d) for $\kappa=1/5$ and $\kappa=5$. Compared to the isotropic disorder case, the depletion decays faster with increasing the correlation length for small value of $\kappa$. In the opposite case, when the radial correlation length is larger than the longitudinal one, the depletion decays much slower with increasing correlation lengths. This can be explained by the fact that, for a fixed value of $z_\rho$, the value of $z_z$ is given by $z_z=z_\rho/\kappa$, which effectively corresponds to a larger disorder correlation length in Fig.~\ref{fig:n0}(b) and leads to a faster decay, while the effective correlation length in Fig.~\ref{fig:n0}(d) is smaller and, hence, the decay is slower. Therefore, we conclude that the ratio of correlation lengths $\kappa$ has a significant impact on the condensate depletion and, thus, can be used for its control. Note that the condensate depletion can be measured in matter interference experiments, where the fragmented part of the fluid contributes with a random phase and, therefore, reduces correspondingly the contrast of the interference pattern.

%%%%%%%%%%%%%%%%%%%%%%%%%%%%%%%%%%%%%%%%%%%%%%%%%%%%%%%%%%%%%%%%%%
%%%%%%%%%%%%%%%%%%%%%%%%%%%%%%%%%%%%%%%%%%%%%%%%%%%%%%%%%%%%%%%%%%
%%%
%%%   chemical potential
%%%
%%%%%%%%%%%%%%%%%%%%%%%%%%%%%%%%%%%%%%%%%%%%%%%%%%%%%%%%%%%%%%%%%%
%%%%%%%%%%%%%%%%%%%%%%%%%%%%%%%%%%%%%%%%%%%%%%%%%%%%%%%%%%%%%%%%%%
%

\subsection{Equation of state}
We now proceed with the perturbative calculation of the chemical potential and the inverse compressibility using Eqs.~(\ref{eq:eqstgen}) and (\ref{eq:dmdngen}). Their dimensionless disorder corrections are defined as
\begin{eqnarray}
\label{eq:deltamudef}
\Delta_\mu &=& \lim_{R\rightarrow0} \frac{\mu-n V({\mathbf k}={\mathbf 0})}{g\,n_{\mathrm HM}}\, ,\\
\label{eq:deltacompressdef}
\Delta_\frac{\partial\mu}{\partial n} &=&\lim_{R\rightarrow0} \frac{\frac{\partial\mu}{\partial n}-V({\mathbf k}={\mathbf 0})}{g\,n_{\mathrm HM}/n}\, ,
\end{eqnarray}
and can be calculated in the similar way as the condensate depletion. 
Analytical results for the general case are given in the Appendix and in the Supplemental Material \cite{supplemental}, while disorder corrections to the chemical potential for the special cases of a pure contact interaction ($\epsilon=0$) and isotropic disorder ($z_\rho=z_z=z$) are given by:
\begin{widetext}
\begin{eqnarray}
\Delta_{\mu}\big|_{\epsilon=0}&=&
-\frac{
	4(z_{\rho}^2-2)
	}{
	(z_{\rho}^2-1)(z_{\rho}+1)(z_{\rho}+z_z)
	}
\, T\left(\frac{z_\rho-1}{z_\rho+1}\, \frac{z_z-z_\rho}{z_z+z_\rho}\right)
+\frac{
	8
	}{
	z_{\rho}+z_z
	}
\, T\left( \frac{z_\rho-z_z}{z_\rho+z_z}\right)\,,\label{eq:eqstsol}\\
\Delta_\mu\big|_{z_\rho=z_z=z}&=&
2\frac{-1+\lambda +2 z \delta  \{-1-\lambda +z \delta  [1-\lambda +z \delta  (1+\lambda )]\}}{z \left(-1+z^2 \delta ^2\right)
[1-\lambda +z \delta  (1+\lambda )]}
+\frac{2 (-1+\lambda )} {z^2 \delta}\, T(-\lambda )\nonumber\\
&&+\frac{2 (-1+\lambda ) }{z^2 \delta  (-1+z \delta
) (1+z \delta )^2}\, T\left(\frac{z \delta-1}{z\delta+1} \lambda \right)\,.\label{eq:eqstsol2}
\end{eqnarray}
\end{widetext}

The analytically calculated disorder corrections to the chemical potential are shown in Fig.~\ref{fig:mu}. The two special cases (\ref{eq:eqstsol}) and (\ref{eq:eqstsol2})
correspond to Figs.~\ref{fig:mu}(a) and \ref{fig:mu}(c), respectively, while Figs.~\ref{fig:mu}(b) and \ref{fig:mu}(d) correspond to the general case with both contact and dipole-dipole interaction as well as anisotropic disorder. The disorder correction increases with increasing disorder strength $R$, regardless of the strength of the dipolar interaction and disorder correlation lengths. This is due to the repulsive interparticle interaction, which has a higher potential energy when the fluid is less uniform. The correction of the chemical potential in the case of the pure contact interaction, shown in Fig.~\ref{fig:mu}(a), has a similar dependence on the correlation lengths as the condensate depletion in Fig.~\ref{fig:n0}(a), while, according to Fig.~\ref{fig:mu}(c), the dipole-dipole interaction does not have a significant effect. This is due to the fact that the dipolar interaction contributes partially as attractive and partially as repulsive, thus leading only to a small net effect. Note that the chemical potential in the clean case is anisotropic, as can be seen from Eq.~(\ref{eq:eqstgen}) and the directional dependence of the limit $\mathbf{k}\rightarrow \mathbf{0}$ in Eq.~(\ref{eq:poten}). This peculiar behaviour is discussed in more detail in Ref.~\cite{lima2012}. For the general case of anisotropic disorder, we see from Figs.~\ref{fig:mu}(b)--\ref{fig:mu}(d) that increasing anisotropy $\kappa=z_\rho/z_z$ leads to a slower decay of the disorder correction with increasing correlation lengths for the same reasons as for the condensate depletion.

\begin{figure*}[!t]
  \includegraphics[width=6cm]{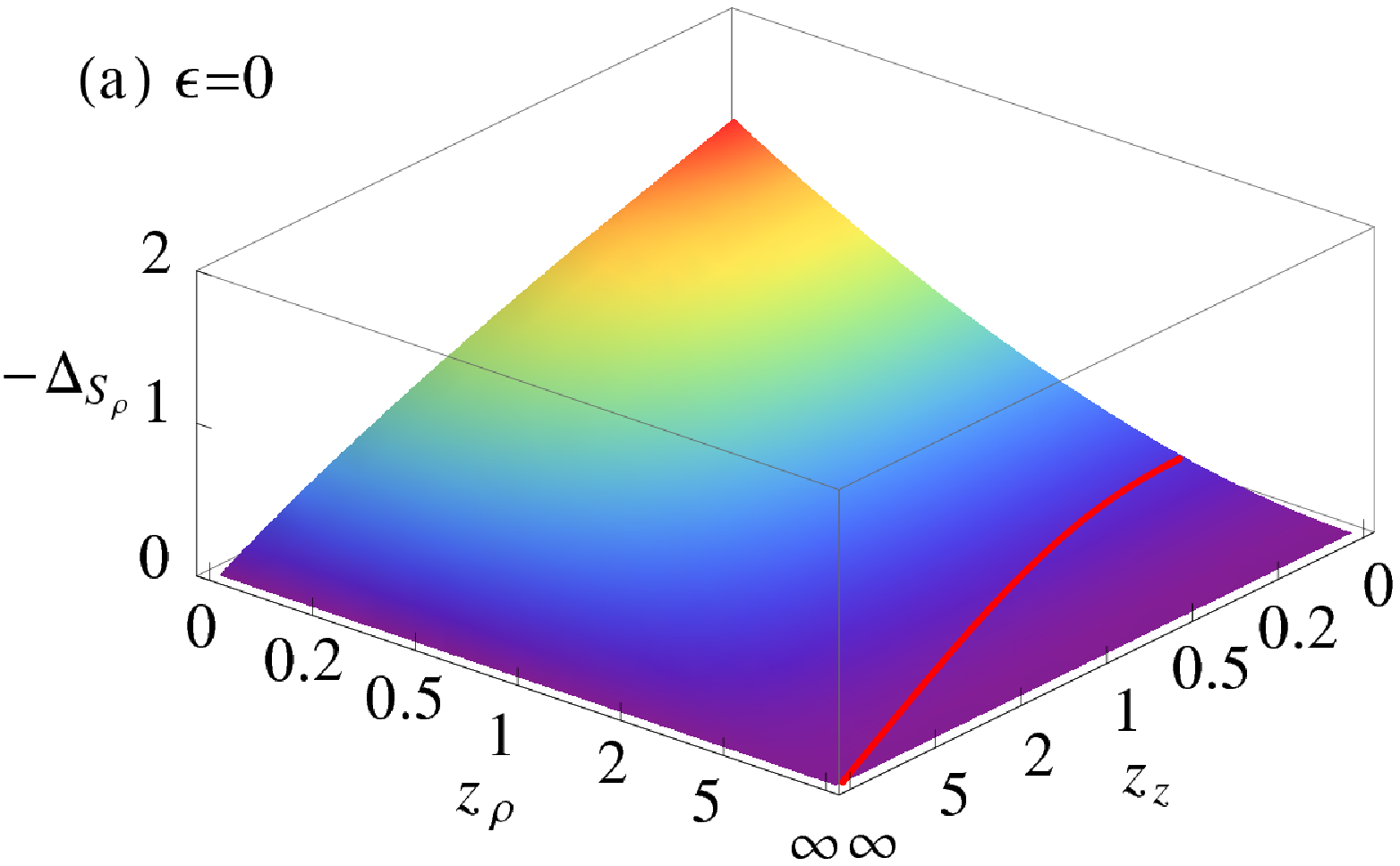}\hspace*{5mm}
  \includegraphics[width=6cm]{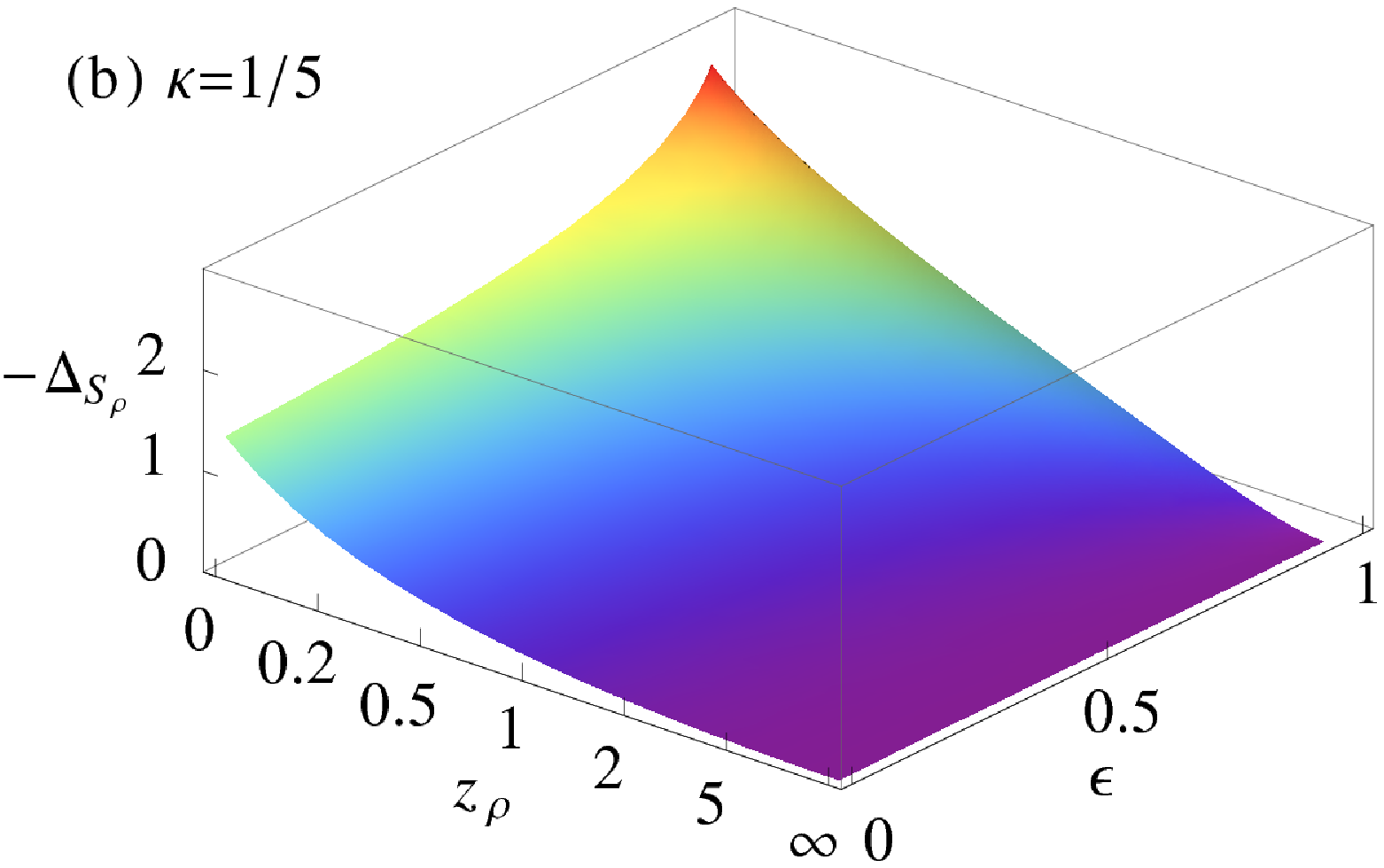}\vspace*{5mm}
  \includegraphics[width=6cm]{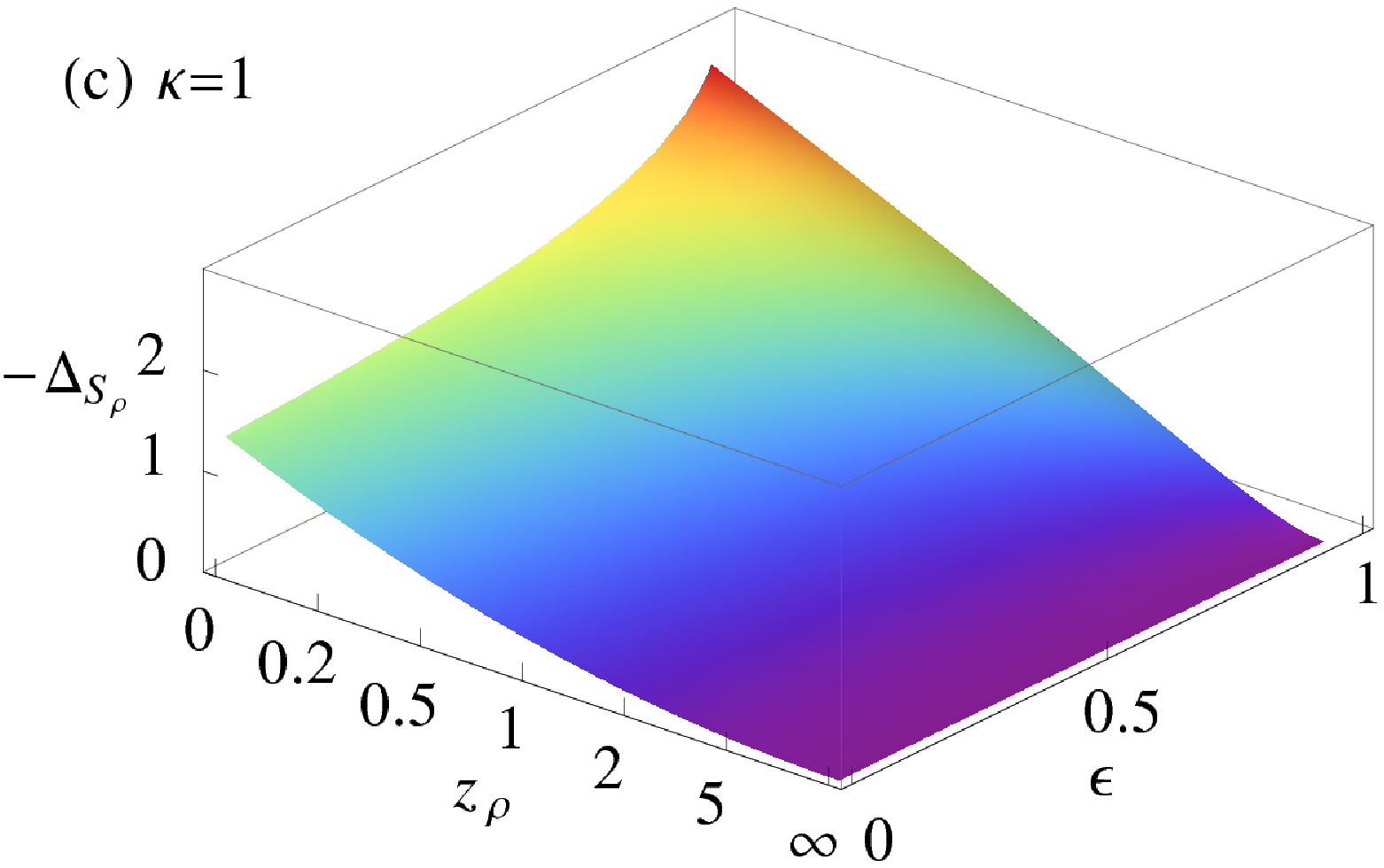}\hspace*{5mm}
  \includegraphics[width=6cm]{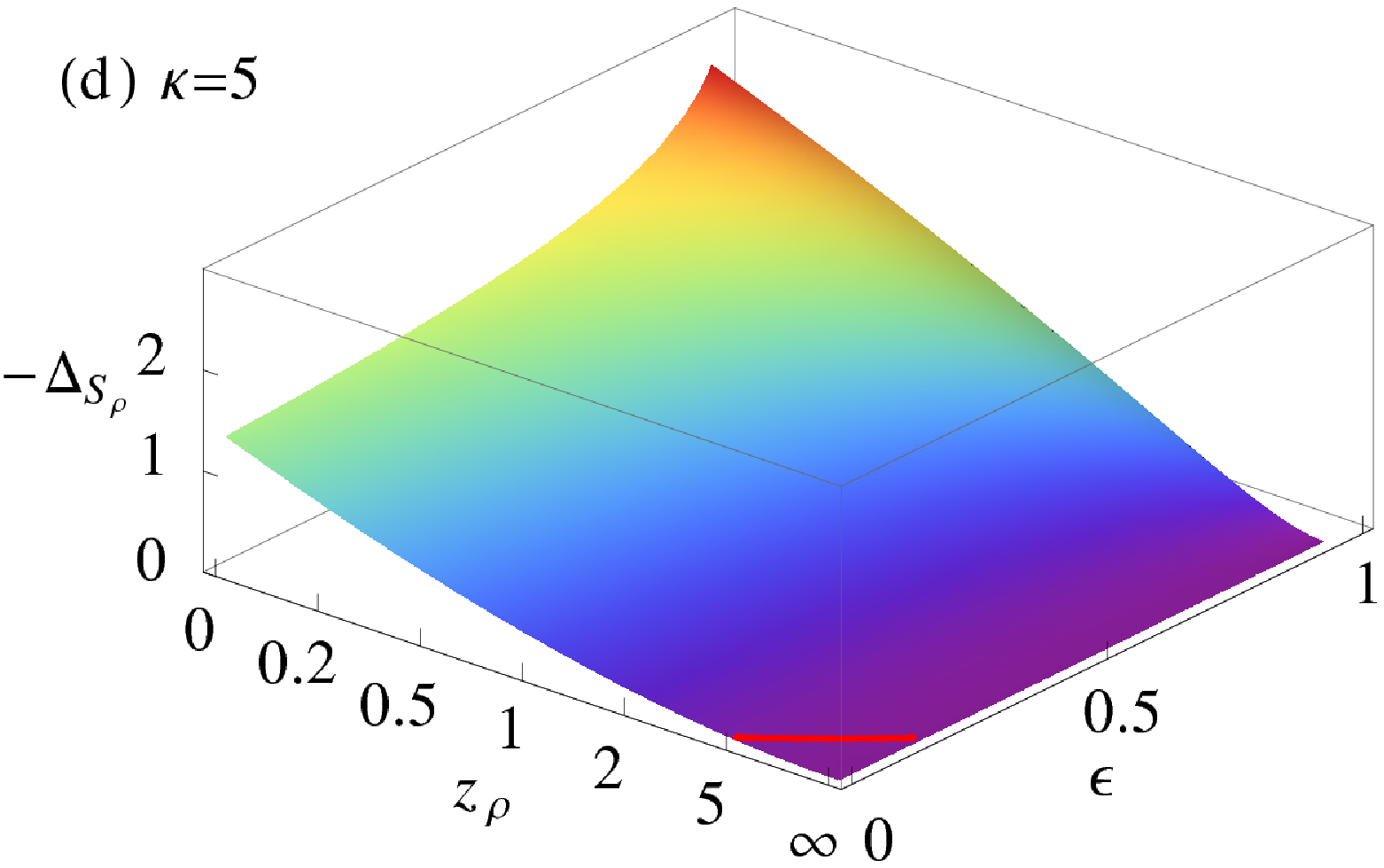}
  \caption{(Color online) Correction to the perpendicular superfluid density due to weak disorder: (a) as a function of the correlation lengths $z_\rho$ and $z_z$ for anisotropic disorder and pure contact interaction ($\epsilon=0$);
(b) as a function of the correlation length $z_\rho$ and the ratio of the dipole-dipole and the contact interaction $\epsilon$ for anisotropic disorder with $\kappa=z_\rho/z_z=1/5$; (c) $\kappa=1$; (d) $\kappa=5$. Red lines in panels (a) and (b) show values of parameters where the perpendicular superfluid depletion becomes equal to the condensate depletion.}
  \label{fig:scperp}
\end{figure*}

\begin{figure*}[!t]
  \includegraphics[width=6cm]{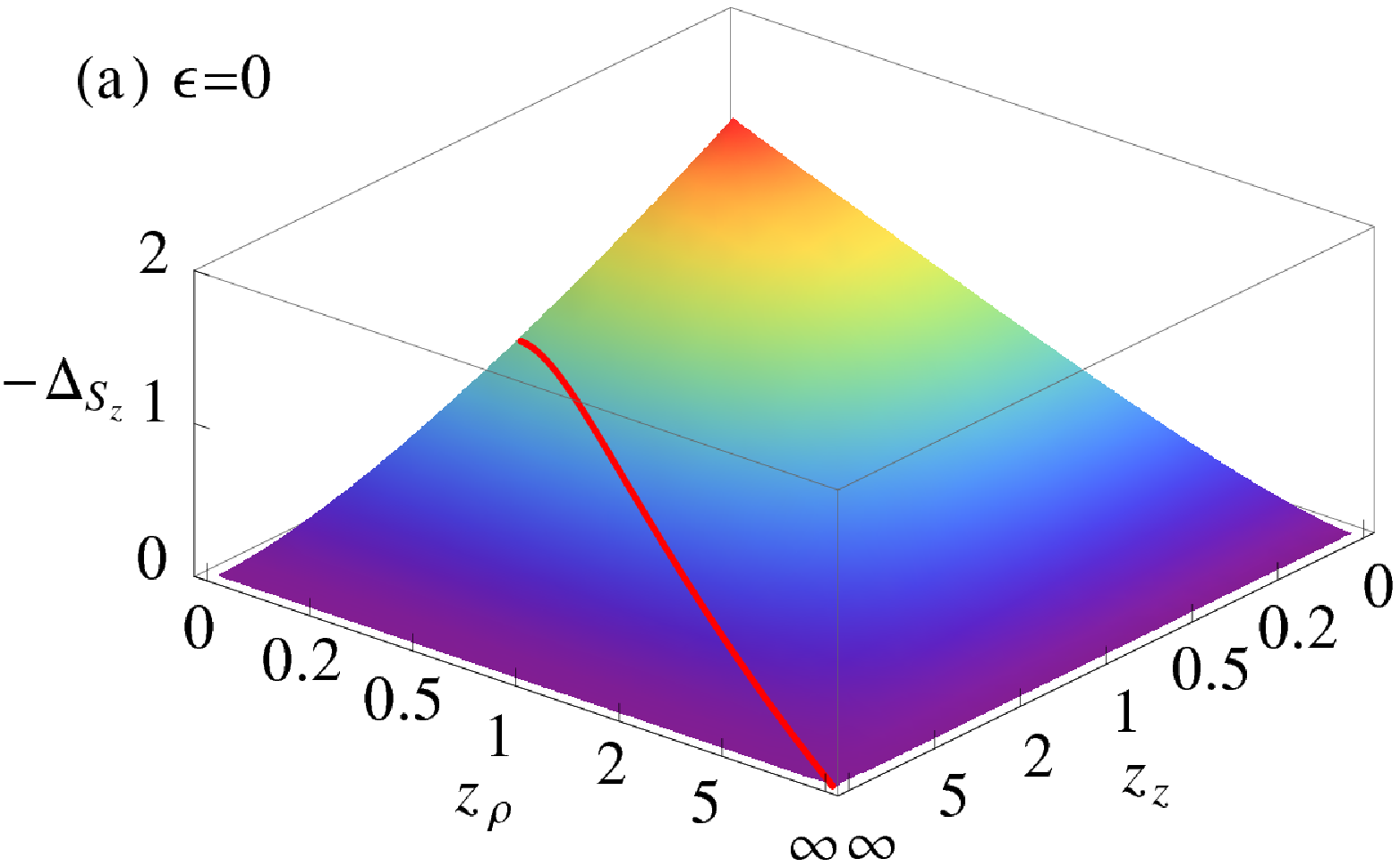}\hspace*{5mm}
  \includegraphics[width=6cm]{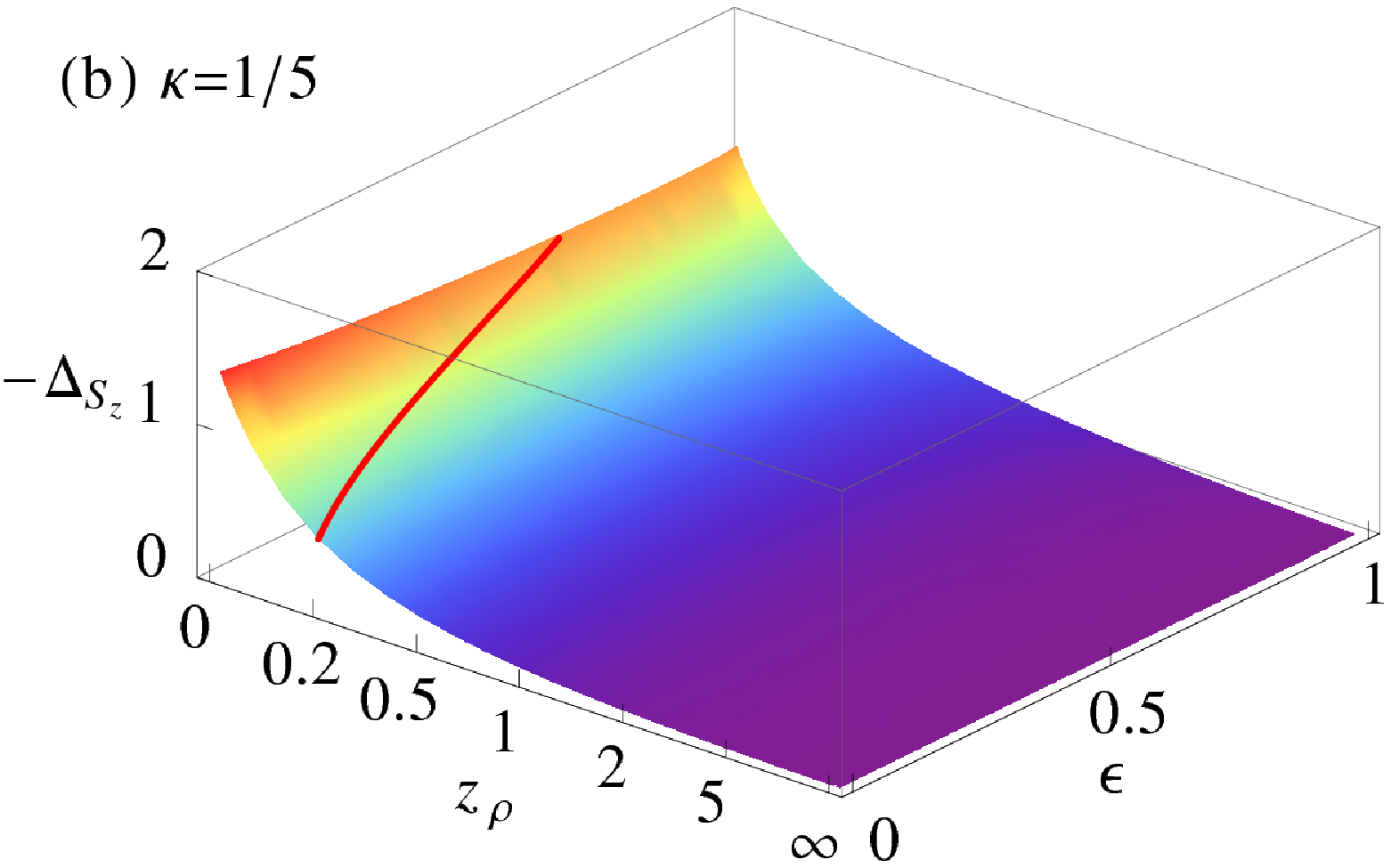}\vspace*{5mm}
  \includegraphics[width=6cm]{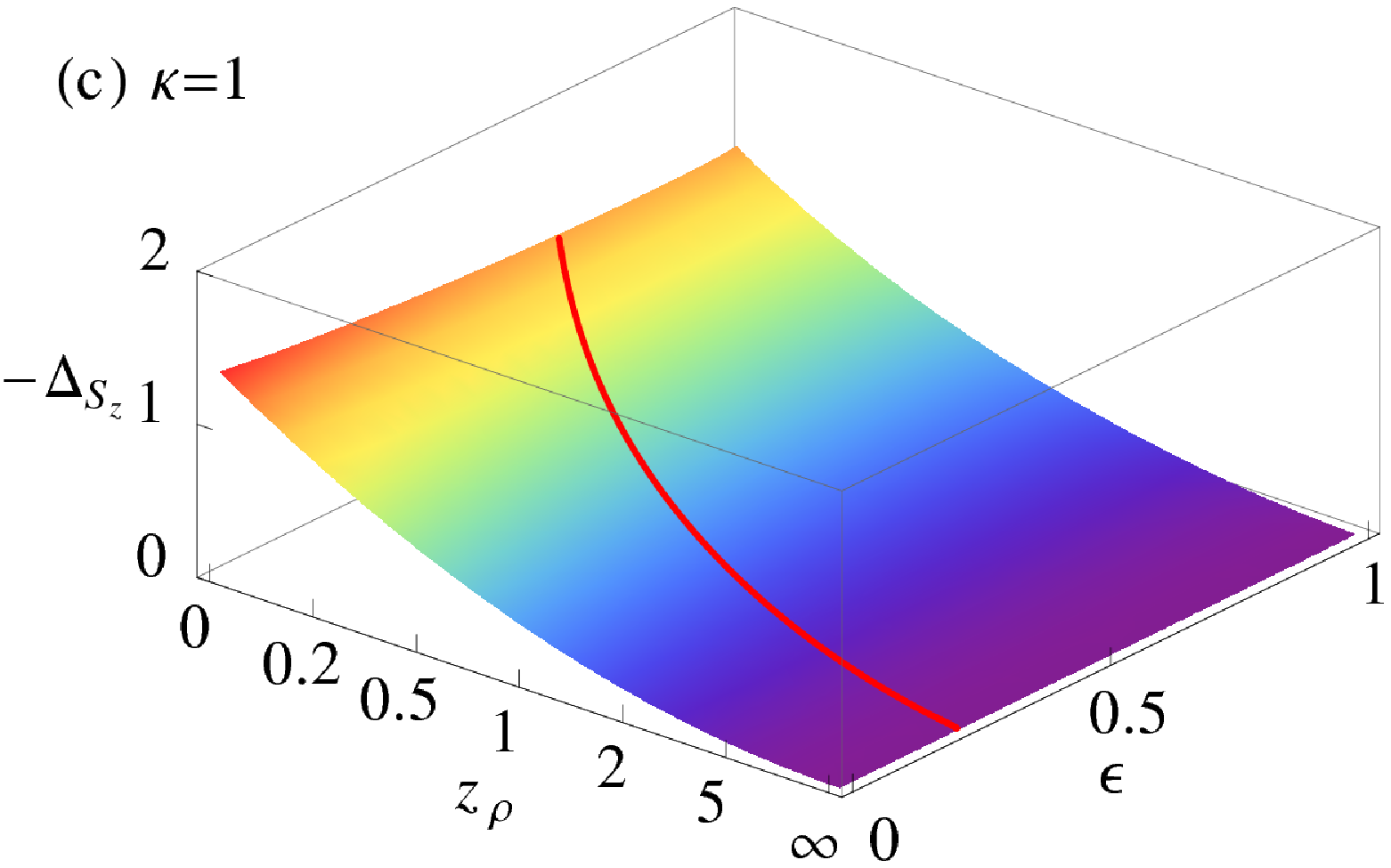}\hspace*{5mm}
  \includegraphics[width=6cm]{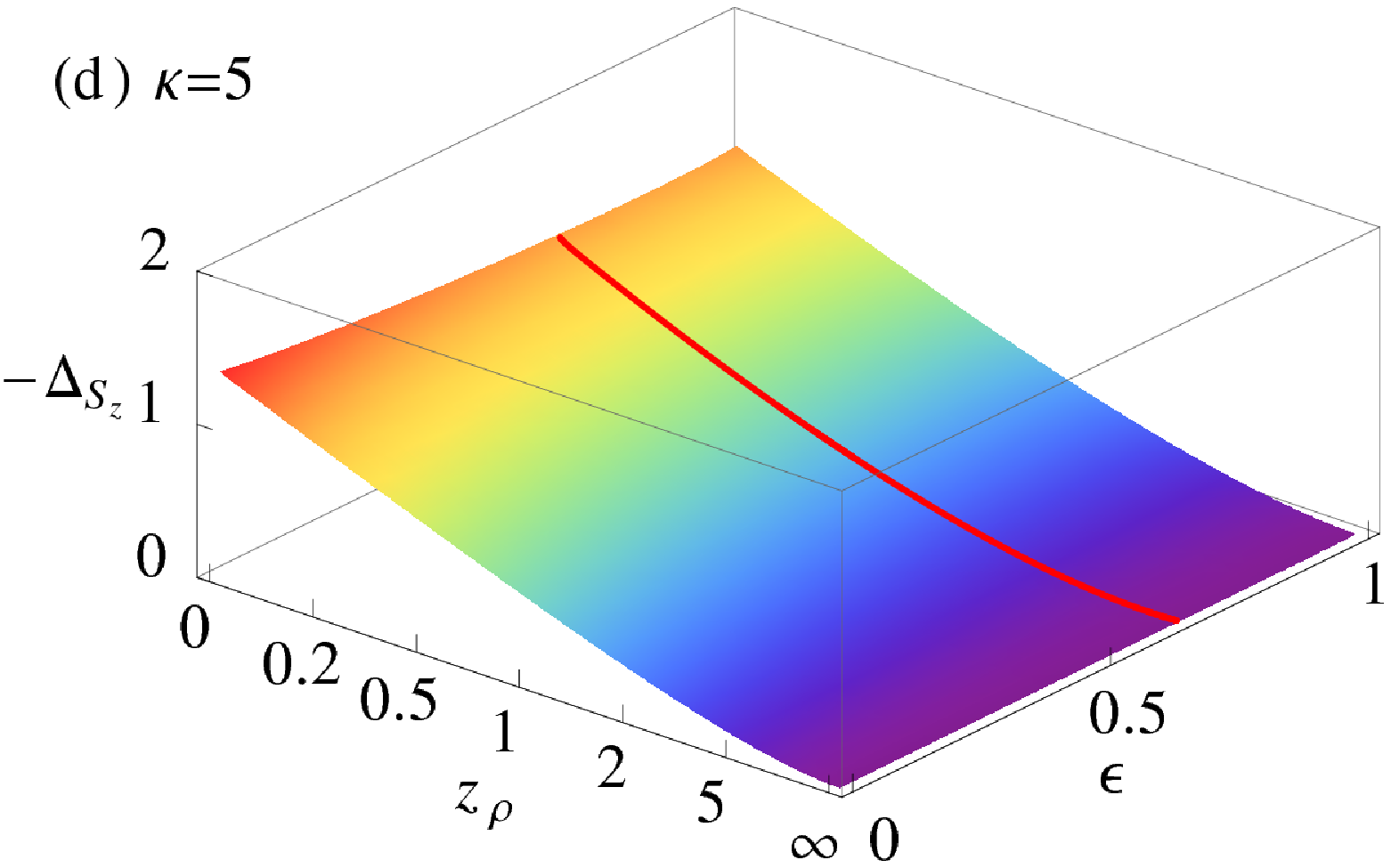}
  \caption{(Color online) Correction to the parallel superfluid density due to weak disorder: (a) as a function of the correlation lengths $z_\rho$ and $z_z$ for anisotropic disorder and pure contact interaction ($\epsilon=0$);
(b) as a function of the correlation length $z_\rho$ and the ratio of the dipole-dipole and the contact interaction $\epsilon$ for anisotropic disorder with $\kappa=z_\rho/z_z=1/5$; (c) $\kappa=1$; (d) $\kappa=5$. Red lines show values of parameters where the parallel superfluid depletion becomes equal to the condensate depletion.}
  \label{fig:scz}
\end{figure*}

The corresponding results for disorder corrections of the inverse compressibility are:
\begin{widetext}
\begin{eqnarray}
\Delta_{\frac{\partial\mu}{\partial n}}\big|_{\epsilon=0}&=&\frac{2}{
\left(-1+z_{\rho }^2\right){}^2}
\left[
\frac{3+z_z \left(2+z_{\rho }^2\right)}{2 \left(1+z_z\right){}^2}+\frac{z_{\rho }^2 \left(-4+z_{\rho }^2\right) }{\left(1+z_{\rho }\right) \left(z_z+z_{\rho }\right)}
T\left(\frac{z_\rho-1}{z_\rho+1}\frac{z_z-z_\rho}{z_z+z_\rho}\right)
\right]\,,\\
\Delta_{\frac{\partial\mu}{\partial n}}\big|_{z_\rho =z_z=z}&=&
\frac{(-1+\lambda ) \left\{1-\lambda +z^2 \delta ^2 [2-2 \lambda +3 z \delta  (1+\lambda )]\right\}}{z \left(-1+z^2 \delta ^2\right)^2
[1-\lambda +z \delta  (1+\lambda )]^2}\,.
\end{eqnarray}
\end{widetext}
They represent intermediate results for calculating later on the sound velocity in subsection \ref{sec:sound}, using Eq.~(\ref{eq:cgen}).

%%%%%%%%%%%%%%%%%%%%%%%%%%%%%%%%%%%%%%%%%%%%%%%%%%%%%%%%%%%%%%%%%%
%%%%%%%%%%%%%%%%%%%%%%%%%%%%%%%%%%%%%%%%%%%%%%%%%%%%%%%%%%%%%%%%%%
%%%
%%%   superfluidity
%%%
%%%%%%%%%%%%%%%%%%%%%%%%%%%%%%%%%%%%%%%%%%%%%%%%%%%%%%%%%%%%%%%%%%
%%%%%%%%%%%%%%%%%%%%%%%%%%%%%%%%%%%%%%%%%%%%%%%%%%%%%%%%%%%%%%%%%%
%

\begin{figure*}[!t]
  \includegraphics[width=6cm]{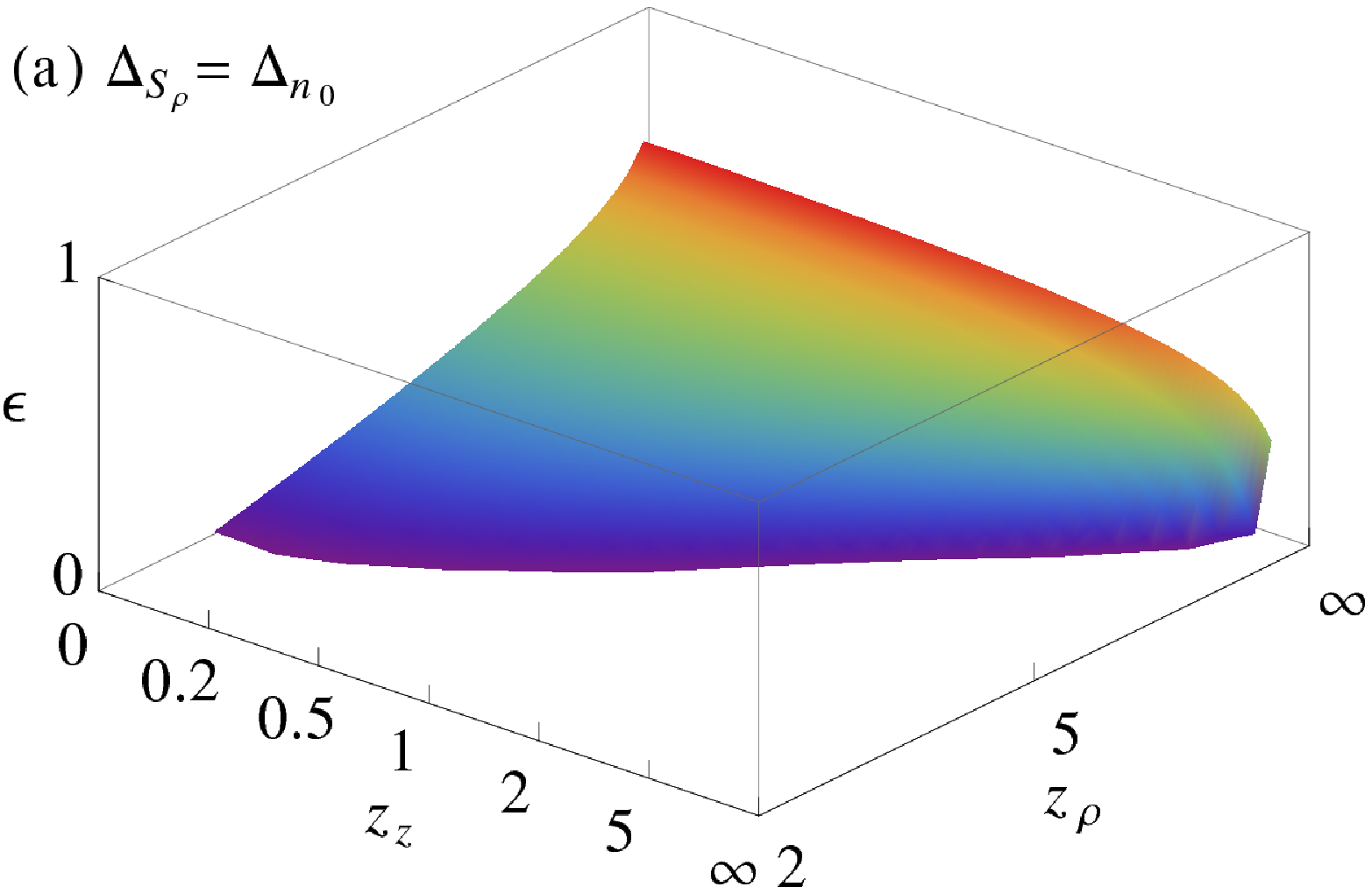}\hspace*{5mm}
  \includegraphics[width=6cm]{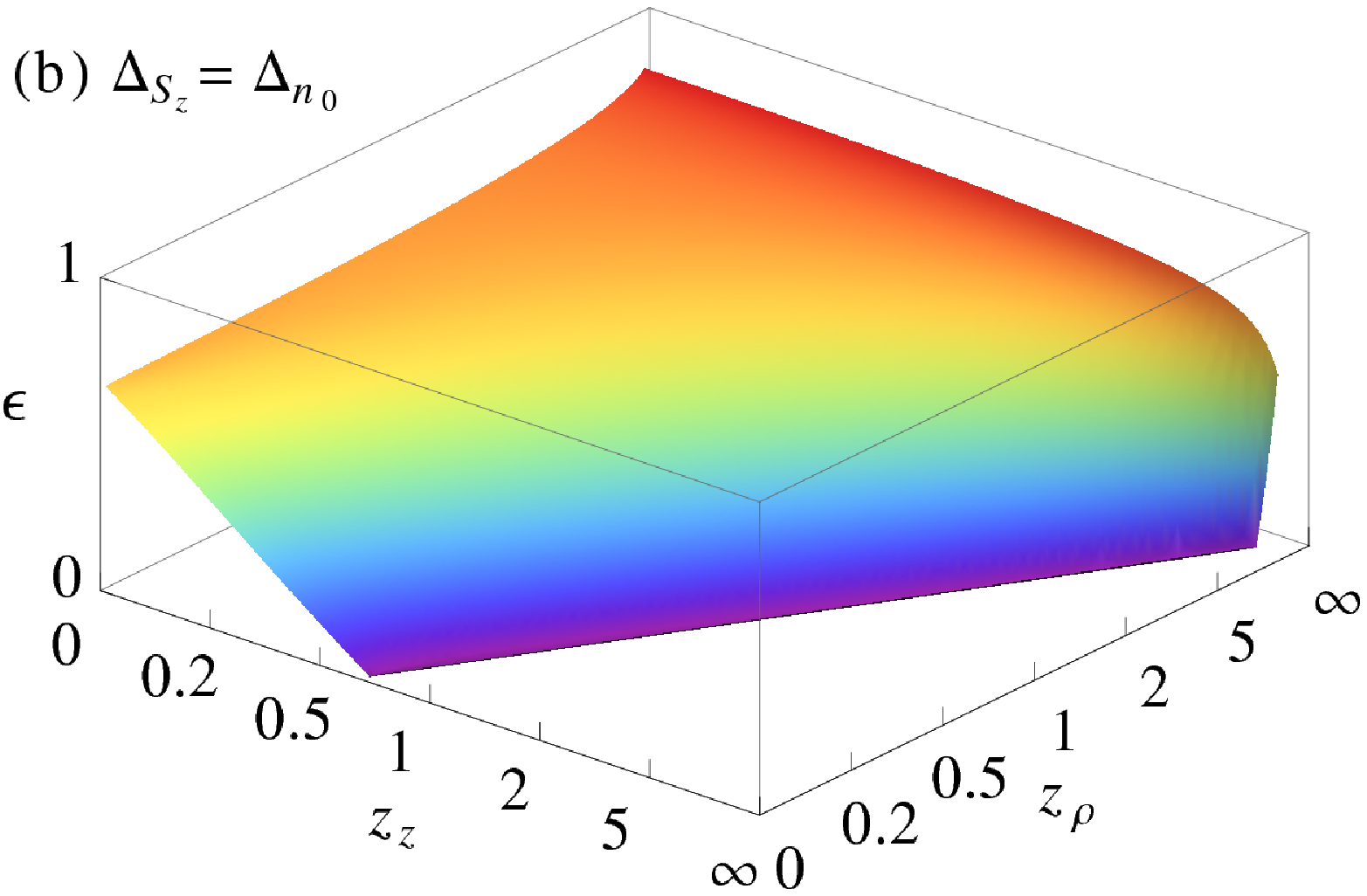}
  \caption{(Color online) Critical values of the ratio of the dipole-dipole and the contact interaction $\epsilon$ at which the superfluid depletion becomes equal to the condensate depletion, as a function of the correlation lengths $z_\rho$ and $z_z$ for: (a) perpendicular and (b) parallel superfluid density.}
  \label{fig:sccrit}
\end{figure*}

\subsection{Superfluidity}
\label{sec:sf}

Now we turn to the calculation of the dimensionless disorder correction to the superfluid density tensor, which is defined by
\begin{equation}
\hat\Delta_{n_S}=\lim_{R\rightarrow0} \frac{\hat{n}_S-n\hat I}{n_{\mathrm HM}}\, .
\end{equation}
It can be separated into a perpendicular and a parallel component after integration with respect to the azimuthal angle $\varphi$. Using Eq.~(\ref{eq:cyl}), we get 
\begin{equation}
\label{eq:sdepcyl}
\hspace*{-1mm}-\hat{\Delta}_{n_S}=
-2\Delta_{n_0}
\left(\begin{array}{ccc}
1  & 0 & 0\\
0 & 1 & 0 \\
0 & 0 & 0
\end{array} \right)
+
2I_{\rm sd}
\left(\begin{array}{ccc}
-1  & 0 & 0\\
0 & -1 & 0 \\
0 & 0 & 2
\end{array} \right)
,
\end{equation}
where $\Delta_{n_0}$ is already calculated in Eqs.~(\ref{eq:depletion}) and (\ref{eq:depletion2}), and $I_{\rm sd}$ is a new integral of the form
\begin{equation}\label{eq:sdcyl}
I_{\rm sd}=\int_0^1 dt\frac{t^2}{v\left(1+v r\right)^2}\, .
\end{equation}
The solution of this integral for the general case is given in the Supplemental Material \cite{supplemental}, while for the two studied special cases the solutions are:
\begin{widetext}
\begin{eqnarray}
&&\hspace*{-10mm}I_{\mathrm{sd}}\big|_{\epsilon=0}=
\frac{1}{\left(-z_z+z_{\rho }\right) \left(z_z+z_{\rho }\right)}\Bigg[-\frac{2+z_z}{1+z_z} +
\frac{2 \left(-2+z_{\rho }^2\right)}{\left(1+z_{\rho }\right)\left(z_z+z_{\rho
}\right)}\,
T\left(\frac{z_\rho-1}{z_\rho+1}\, \frac{z_z-z_\rho}{z_z+z_\rho}\right)+
\frac{4}{z_z+z_{\rho }}\,
 T\left(\frac{z_{\rho }-z_z}{z_{\rho }+z_z}\right)
\Bigg]\,,\\
&&\hspace*{-10mm}I_{\rm sd}\big|_{z_\rho =z_z=z}=\frac{(-1+\lambda )^3}{4 z \delta ^2 \lambda  [1-\lambda +z \delta  (1+\lambda )]}
-\frac{(-1+\lambda
)^3 }{4 z^2 \delta ^3 \lambda }\, T(-\lambda )+
\frac{(-1+\lambda )^3 }{4 z^2 \delta ^3 (\lambda +z \delta  \lambda )}\, T\left(\frac{z \delta-1}{z\delta+1} \lambda \right)\,.
\end{eqnarray}
\end{widetext}

In an arbitrary direction of an unit vector $\mathbf{q}$ the superfluid density can be calculated by describing the tensorial superfluid density according to $n_S(\mathbf{q})=\mathbf{q}\hat n_S\mathbf{q}$. In the case of cylindrical symmetry this reduces to
\begin{equation}\label{eq:sup_Separation}
n_S(\mathbf{q})=n_{S_\rho}\sin^2\phi(\mathbf{q},\mathbf{e}_z)+n_{S_z}\cos^2\phi(\mathbf{q},\mathbf{e}_z)\, .
\end{equation}
Thus, obtaining the disorder corrections for $n_{S_\rho}$ and $n_{S_z}$ is sufficient for recovering the superfluid depletion in any direction. From Eq.~(\ref{eq:sdepcyl}) we directly read off
\begin{eqnarray}
-\Delta_{S_\rho}&=&-2\Delta_{n_0}-2I_{\rm sd}\,,\\
-\Delta_{S_z}&=&4I_{\rm sd}\,.
\end{eqnarray}
Note that the negative sign in front of $\Delta_{S_{\rho,z}}$ and $\Delta_{n_0}$ suggests that the changes of the superfluid densities and the condensate density are negative or, equivalently, that the depletion is positive.
For isotropic systems $\Delta_{S_\rho}$ and $\Delta_{S_z}$ both are equal to $\frac{4}{3}\Delta_{n_0}$, as can be seen from Eq.~(\ref{eq:isot}). Due to an anisotropy, however, there is a range of correlation lengths and relative dipolar interaction strengths, where the  superfluid depletion is smaller than the condensate depletion. Some particles from the fragmented fluid contribute to superfluidity, suggesting that they are not localized indefinitely, but have some finite localization time. This localization time 
was exemplarily calculated in Ref.~\cite{graham2009} within the Hartree-Fock theory of dirty bosons with delta-correlated disorder.

The superfluid depletion in the case of a pure contact interaction and anisotropic disorder shows a similar behavior as the condensate depletion, as can be seen in Figs.~\ref{fig:scperp}(a) and \ref{fig:scz}(a). 
In the presence of disorder as well as both contact and dipole-dipole interaction, Figs.~\ref{fig:scperp}(b)-\ref{fig:scperp}(d) show that the depletion of the perpendicular component is similar to the condensate depletion, but the depletion of the parallel component decreases as the relative interaction strength increases, as is depicted in Figs.~\ref{fig:scz}(b)--\ref{fig:scz}(d).
The red lines in Figs.~\ref{fig:scperp} and \ref{fig:scz} show where the corresponding superfluid depletion component becomes equal to the condensate depletion. This is illustrated in more detail in Fig.~\ref{fig:sccrit}, which presents the values of the interaction ratio $\epsilon$ for which the superfluid depletions become equal to the condensate depletion.

Although defined only for systems without a trap, the above calculated superfluid density could be extended to the trapped case in the simplest way by assuming that it depends only on the local density. If we turn on a slowly moving disorder for a short time $\tau$, such that $\mathbf{v} \tau$ is much smaller than the size of the trap, before switching off the trap, this would change the momentum distribution which could be afterwards reconstructed from a time-of-flight measurement.
In this way our predictions for the superfluid density in such a system might become observable in experiment.

%%%%%%%%%%%%%%%%%%%%%%%%%%%%%%%%%%%%%%%%%%%%%%%%%%%%%%%%%%%%%%%%%%
%%%%%%%%%%%%%%%%%%%%%%%%%%%%%%%%%%%%%%%%%%%%%%%%%%%%%%%%%%%%%%%%%%
%%%
%%%   sound velocity
%%%
%%%%%%%%%%%%%%%%%%%%%%%%%%%%%%%%%%%%%%%%%%%%%%%%%%%%%%%%%%%%%%%%%%
%%%%%%%%%%%%%%%%%%%%%%%%%%%%%%%%%%%%%%%%%%%%%%%%%%%%%%%%%%%%%%%%%%

\begin{figure*}[!t]
  \includegraphics[width=6cm]{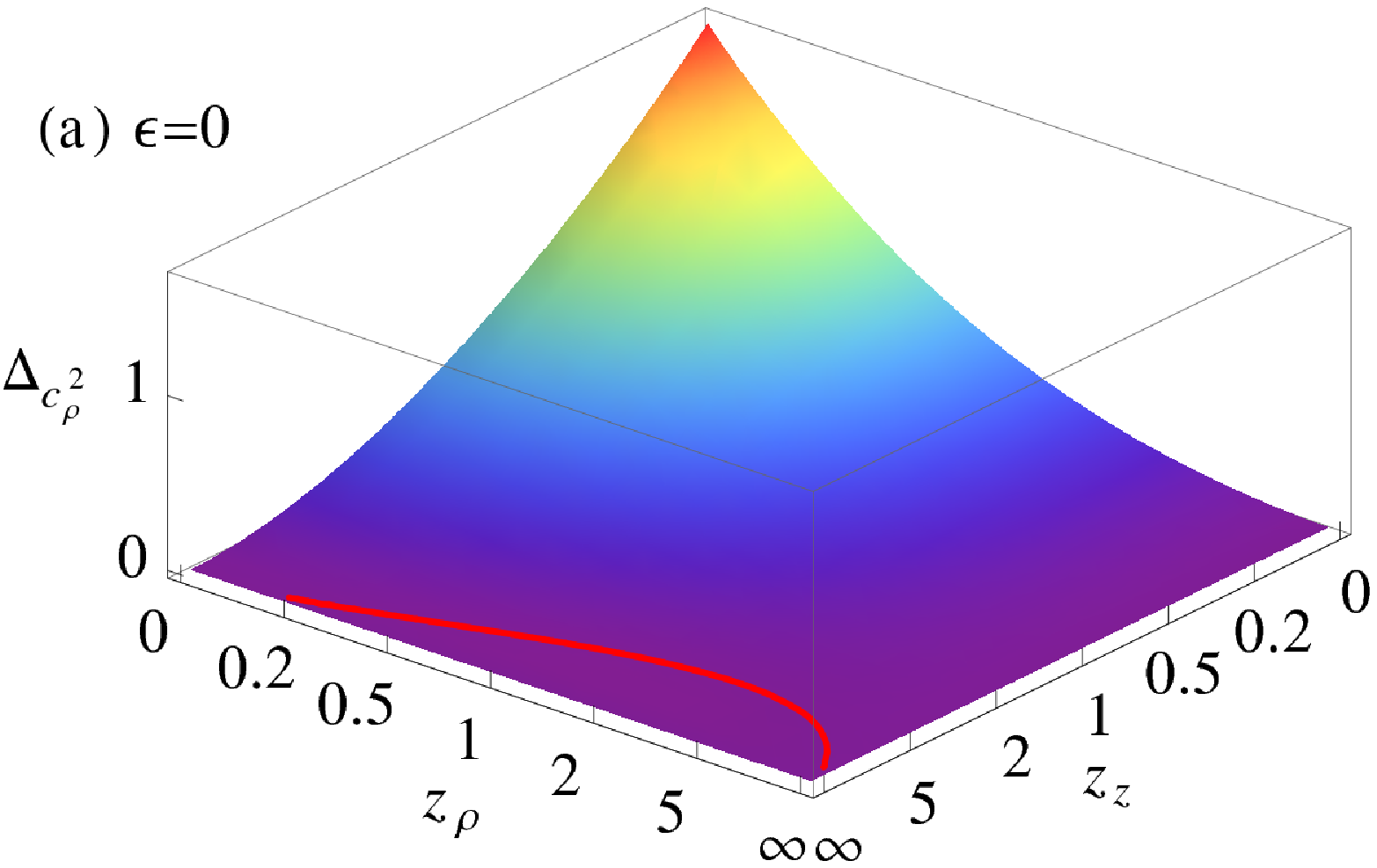}\hspace*{5mm}
  \includegraphics[width=6cm]{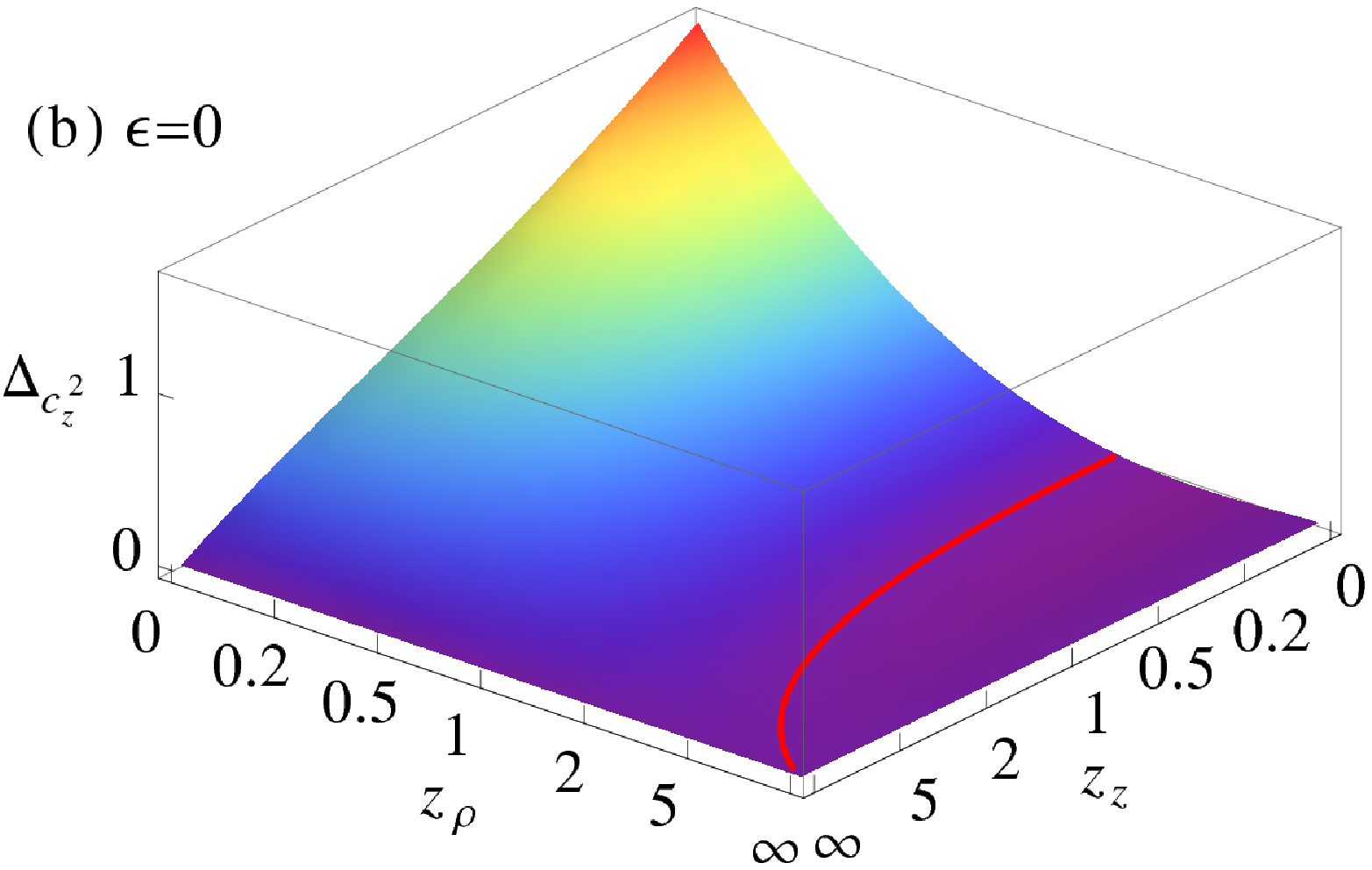}
  \caption{(Color online) Corrections to (a) perpendicular and (b) parallel sound velocity for weak anisotropic disorder and pure contact interaction ($\epsilon=0$) as a function of the correlation lengths $z_\rho$ and $z_z$. The red lines show values of the correlation lengths for which the correction vanishes.}
  \label{fig:soundcont}
\end{figure*}

\begin{figure*}[!t]
  \includegraphics[width=6cm]{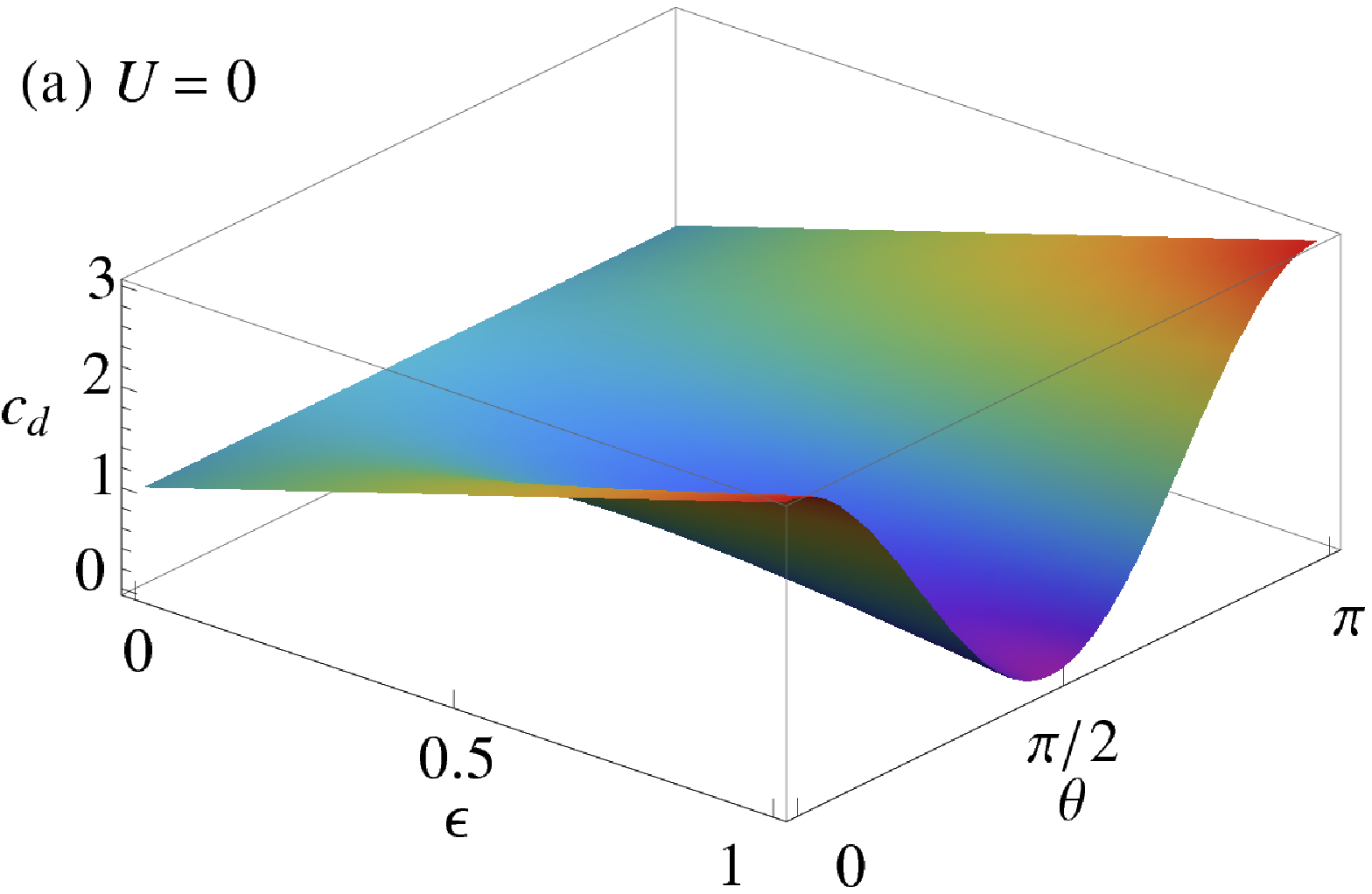}\hspace*{5mm}
  \includegraphics[width=6cm]{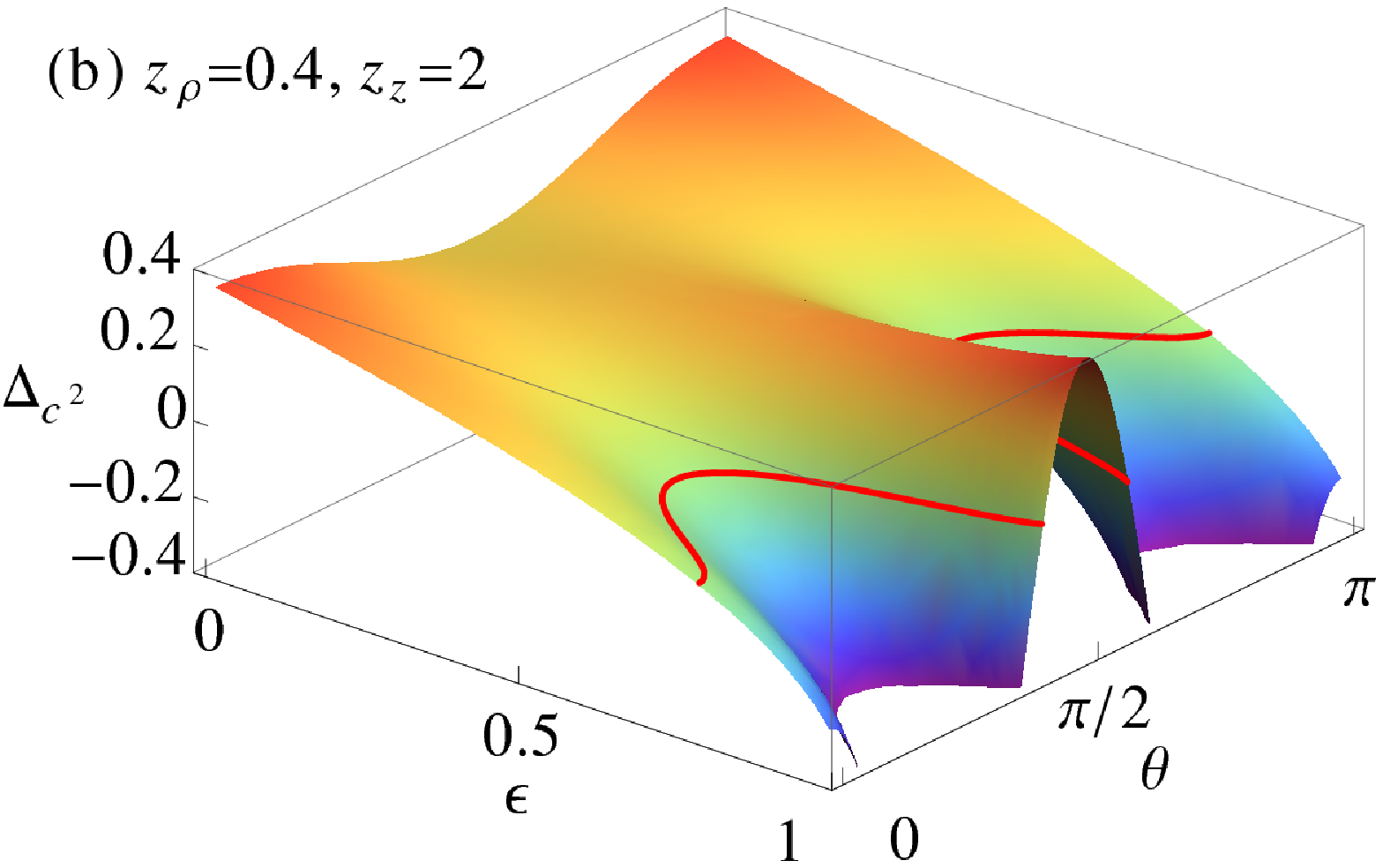}\vspace*{1mm}
  \includegraphics[width=6cm]{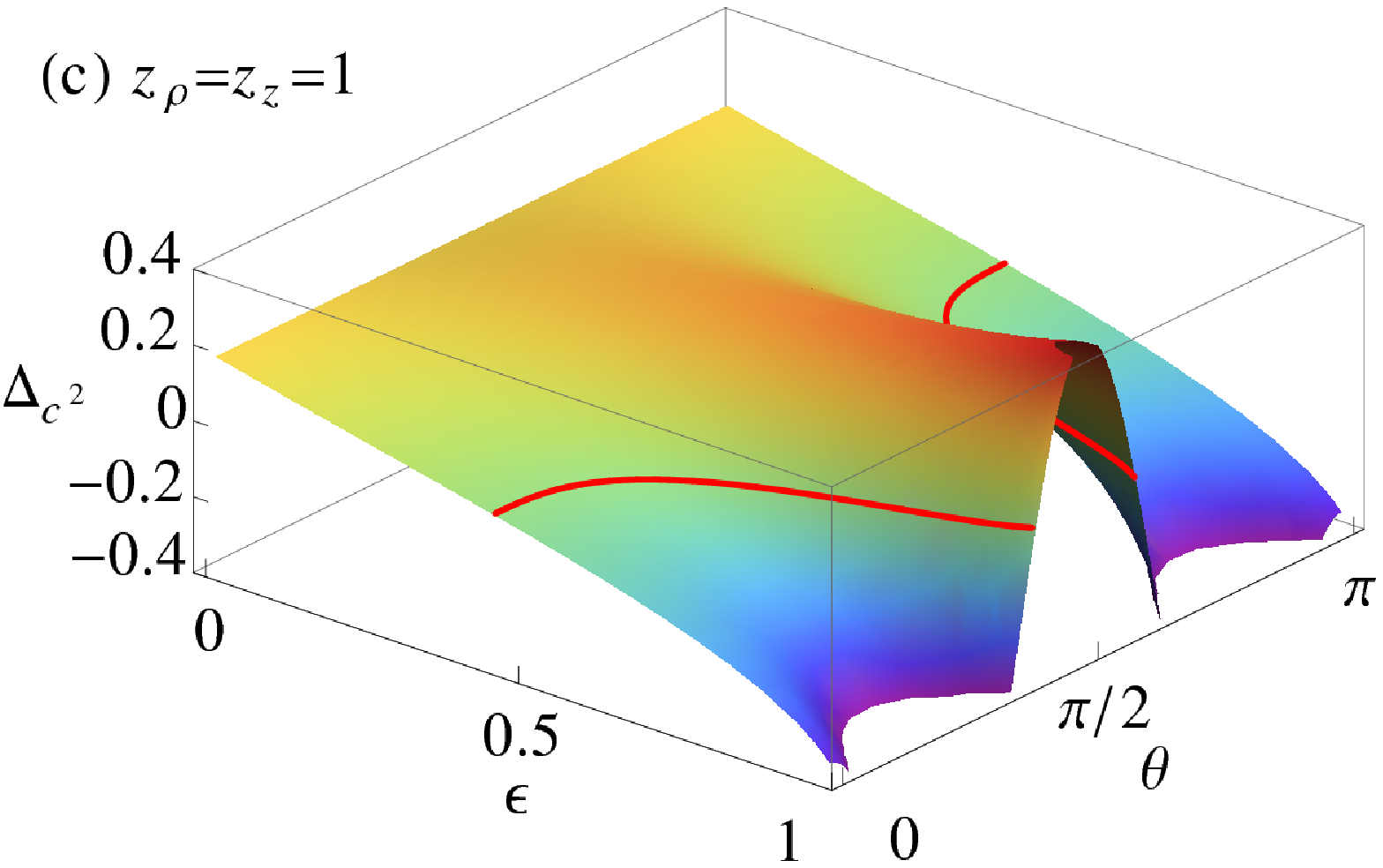}\hspace*{5mm}
  \includegraphics[width=6cm]{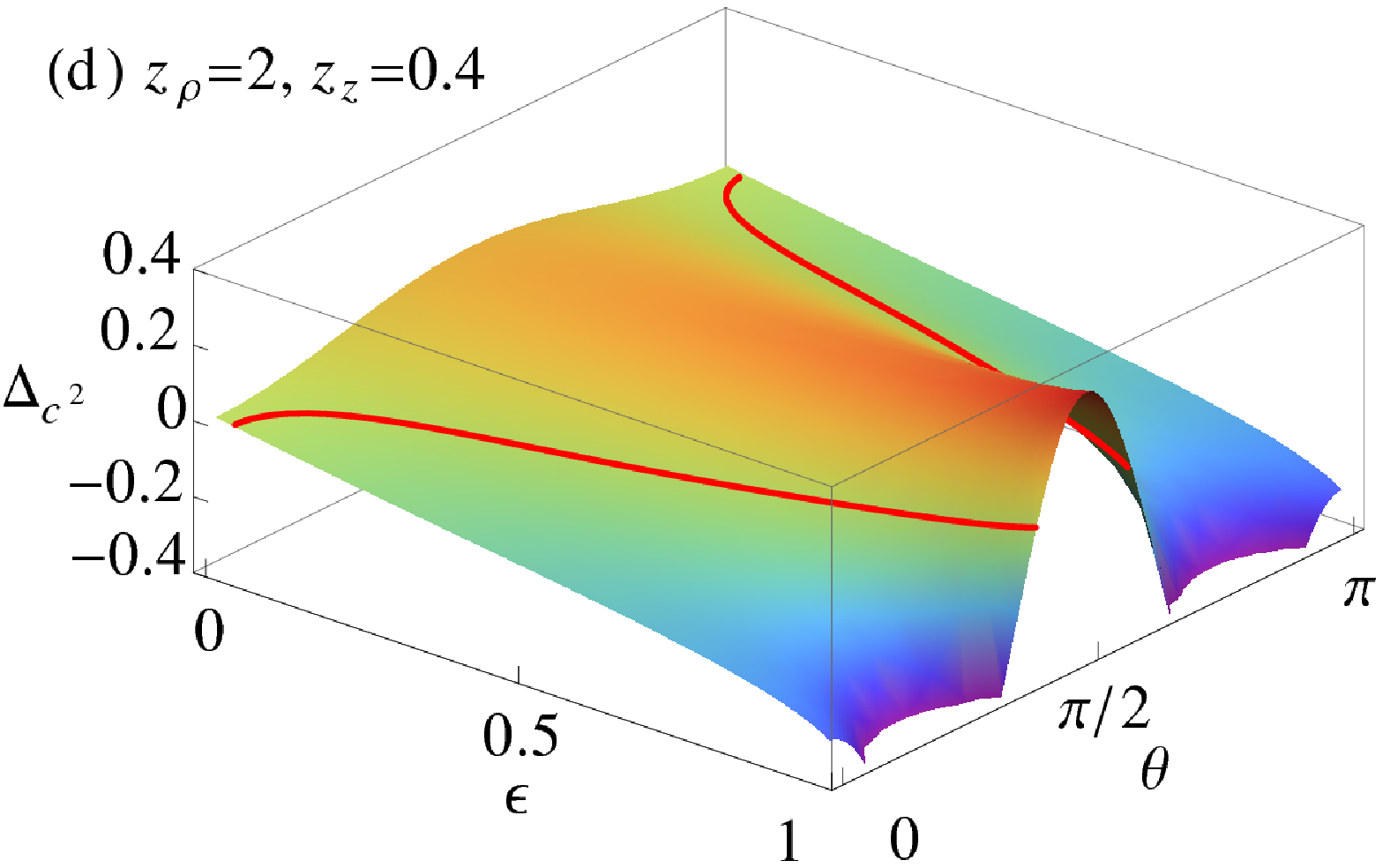}
  \caption{(Color online) (a) Sound velocity for the clean case (no disorder) as a function of the ratio of the dipole-dipole and the contact interaction $\epsilon$ and the azimuthal angle $\theta=\phi(\mathbf{q}, \mathbf{e}_z)$; (b) correction to the sound velocity due to weak delta--correlated disorder as a function of the ratio of the dipole-dipole and the contact interaction $\epsilon$ and the azimuthal angle $\theta=\phi(\mathbf{q}, \mathbf{e}_z)$ for $z_\rho=0.4$, $z_z=2$; (c) $z_\rho=z_z=1$; (d) $z_\rho=2$, $z_z=0.4$. The red lines show values of the parameters $\epsilon$ and $\theta$ for which the correction vanishes.}
  \label{fig:sound}
\end{figure*}

\subsection{Sound velocity}
\label{sec:sound}
The corresponding disorder correction of the sound velocity in the direction of the unit vector $\mathbf{q}$ can be calculated using Eq.~(\ref{eq:cgen}):
\begin{eqnarray}
\hspace*{-5mm}c_{\mathbf{q}}^2&=&\frac{g n}{m}\left[
\frac{V_{\mathbf{q}}(\mathbf{0})}{g}\right.\\
&&\left. + \frac{n_{\mathrm{HM}}}{n}\left(\frac{V_{\mathbf{q}}(\mathbf{0})}{g}\mathbf{q}^T \hat{\Delta}_{n_S}\mathbf{q}+
\Delta_{\frac{\partial\mu}{\partial n}}\right)+\ldots
\right]\,,\nonumber
\end{eqnarray}
If we define the dimensionless disorder correction as
\begin{equation}
\Delta_{c^2_\mathbf{q}}=\lim_{R\rightarrow0} \frac{{c^2_\mathbf{q}}-n V({\mathbf k}={\mathbf 0})/m}{g\,n_{\mathrm HM}/m}\, ,
\end{equation}
from the previous equation we get
\begin{equation}
\Delta_{c_{\mathbf{q}}^2}=\frac{V_{\mathbf{q}}(\mathbf{0})}{g}\mathbf{q}^T \hat{\Delta}_{n_S}\mathbf{q}+
\Delta_{\frac{\partial\mu}{\partial n}}\,,
\end{equation}
where $V_{\mathbf{q}}(\mathbf{0})=\lim_{k\rightarrow 0} V(k \mathbf{q})$ denotes the directional dependence of the potential $V$ on $\mathbf{q}$, according to Eq.~(\ref{eq:poten}).

The anisotropy of disorder comes into play in a simple way. From Eqs.~(\ref{eq:cgen}) and (\ref{eq:sup_Separation}) it follows that the sound velocity can also be separated into a parallel and a perpendicular component
\begin{eqnarray}\label{eq:sound_Separation}
c^2(\mathbf{q})=c^2_\rho\sin^2\phi(\mathbf{q},\mathbf{e}_z)+c^2_z\cos^2\phi(\mathbf{q},\mathbf{e}_z)\,,
\end{eqnarray}
with
\begin{eqnarray}
c^2_{\rho,z}=\frac{1}{m}\frac{\partial \mu}{\partial n} n_{S_{\rho,z}}\;
\end{eqnarray}
and the corresponding dimensionless disorder correction
\begin{eqnarray}
\Delta_{c^2_{\rho,z}}=\Delta_{\frac{\partial\mu}{\partial n}}
    +\Delta_{S_{\rho,z}}\,.
\end{eqnarray}
Figure~\ref{fig:soundcont} shows disorder corrections to the perpendicular and the parallel component of the sound velocity for a pure contact interaction. The red lines correspond to the values of correlation lengths for which the disorder corrections vanish.

For the general case, the anisotropy factor due to the dipolar interaction $c_d=V_{\mathbf{q}}(\mathbf{0})/g$ \cite{krumnow2011} is plotted in Fig.~\ref{fig:sound}(a). By introducing weak disorder, the sound velocity changes via two competing effects: the decrease of the compressibility, i.e. the increase in the inverse compressibility, from Eq.~(\ref{eq:dmdngen}), which tends to increase the sound velocity, and the decrease of the superfluid density corresponding to a negative value of $\Delta_{S_{\rho,z}}$, which tends to decrease the sound velocity. The corresponding results are shown in Fig.~\ref{fig:sound}(b)--\ref{fig:sound}(d). For small disorder correlation lengths the decrease in compressibility is dominant.
These corrections can be experimentally measured, for instance, by determining the phonon dispersion relation by using Bragg spectroscopy \cite{davis1995,stenger1999,ozeri2005}. 

\section{Conclusions}
\label{sec:con}

We have analyzed in detail how the anisotropy of both the dipolar interaction and the presence of disorder affects the directional dependence of different physical observables of dirty Bose-Einstein condensates. Using the mean-field approach at zero temperature, we have calculated the condensate depletion due to disorder, as well as the corresponding corrections to the equation of state, the superfluid density, and the sound velocity. In particular, we have discussed the consequences for the superfluid density, which becomes a tensorial quantity as a linear response to the moving disorder. Whereas Ref.~\cite{krumnow2011} analyzed a dipolar BEC in isotropic disorder potential, we have shown here that the anisotropic disorder provides a separate origin for the tensorial nature of the superfluid density. We have found that a large enough disorder anisotropy can even make both the parallel and perpendicular superfluid density component larger than the corresponding condensate density, which happens in the case of dipolar interaction and isotropic disorder only for the parallel component  \cite{krumnow2011}. 

These initial results necessitate further studies, as they contribute to the overall physical picture in which the localization of bosons in the respective minima of the disorder potential occurs at a characteristic time scale \cite{graham2009}. This localization time remains to be analyzed in more detail in a forthcoming publication. We also plan to study effects of disorder on nonlinear oscillation modes and Faraday waves in BEC \cite{nicolin2011,balaz2012,nicolin2012}.

\section*{Acknowledgements}
We would like to thank Bakhodir Abdullaev, Christian Krumnow, Aristeu R.~P.~Lima, and Vladimir Lukovi\' c for fruitful discussions.
This work was supported in part by the Ministry of Education, Science, and Technological Development of the
Republic of Serbia under projects No.~ON171017 and NAI-DBEC, by DAAD - German
Academic and Exchange Service under project NAI-DBEC, and by the European
Commission under EU FP7 projects PRACE-2IP, PRACE-3IP, HP-SEE, and EGI-InSPIRE.
Both A.~B. and A.~P. gratefully acknowledge a fellowship from the Hanse-Wissenschaftskolleg.

\appendix*
\section{Expressions for $\Delta_{n_0}$, $\Delta_\mu$ in general case}
\label{sec:app}
This Appendix contains expressions for the dimensionless condensate depletion $\Delta_{n_0}$ due to disorder and the dimensionless disorder correction to the chemical potential $\Delta_\mu$ in the general case, when both contact and dipole-dipole interaction are present and characterized by the ratio $\epsilon=C_\mathrm{dd}/3g$. We assume that disorder is cylindrically-symmetric and Lorentzian-correlated, characterized by the dimensionless correlation lengths $z_\rho$ and $z_z$. These expressions, as well as the expressions for the integral $I_\mathrm{sd}$ and the dimensionless disorder correction to the inverse compressibility $\Delta_\frac{\partial\mu}{\partial n}$, which define the dimensionless disorder corrections to the superfluid density and the sound velocity, are given in the Supplemental Material \cite{supplemental}, as a {\it Mathematica} notebook.

The expressions for disorder corrections are given in terms of auxiliary functions:
\begin{eqnarray}
T(x)&=&\frac{\arctan\sqrt{x}}{\sqrt{x}}\, ,\\
A(x, y)&=&T(x+\sqrt{y})+T(x-\sqrt{y})\, ,\\
B(x,y)&=& \sqrt{y}\, \left[T(x+\sqrt{y})-T(x-\sqrt{y})\right]\, ,
\end{eqnarray}
and their values
$A_i=A(x_i, y_i)$, $B_i=B(x_i, y_i)$, where arguments $x_i$, $y_i$ ($i=1,2$) are given by:
\begin{widetext}
\begin{eqnarray}
&&x_1=\frac{2 z_{\rho }^2-(1+2 \epsilon ) z_{\rho }^4+z_z^2 \left[-2-(-1+\epsilon ) z_{\rho }^2\right]}{2 z_z^2 \left[1+(-1+\epsilon ) z_{\rho
}^2\right]}\, ,\quad 
x_2=\frac{-6 \epsilon -(-1+\epsilon )^2 z_z^2+\left(1+\epsilon -2 \epsilon ^2\right) z_{\rho }^2}{2 (1+2 \epsilon ) \left[1+(-1+\epsilon ) z_{\rho
}^2\right]}\, ,\\
&&y_1=\frac{z_z^2 \left[12 \epsilon +(-1+\epsilon )^2 z_z^2\right] z_{\rho }^4+2 \left[-6 \epsilon +(-1+\epsilon ) (1+2 \epsilon ) z_z^2\right]
z_{\rho }^6+(1+2 \epsilon )^2 z_{\rho }^8}{4 z_z^4 \left[1+(-1+\epsilon ) z_{\rho }^2\right]{}^2}\, ,\nonumber \\
&&y_2=\frac{(-1+\epsilon )^2 \left(\sqrt{1-\epsilon }-\sqrt{1+2 \epsilon }\right)^4 \left\{(-1+\epsilon )^2 z_z^4-12 \epsilon  z_{\rho }^2+(1+2
\epsilon )^2 z_{\rho }^4+2 z_z^2 \left[6 \epsilon +(-1+\epsilon ) (1+2 \epsilon ) z_{\rho }^2\right]\right\}}{4 \left(-1-2 \epsilon +\sqrt{1+\epsilon
-2 \epsilon ^2}\right)^4 \left[1+(-1+\epsilon ) z_{\rho }^2\right]{}^2}\, .\nonumber
\end{eqnarray}

The dimensionless condensate depletion $\Delta_{n_0}$ due to disorder, defined by Eq.~(\ref{eq:deltan0def}), is given by:
\begin{eqnarray}
\label{eq:appdeln0}
&&\hspace*{-3mm}\Delta_{n_0} = A_1\, \frac{ \left\{-z_z^2 \left[18 \epsilon +(-1+\epsilon )^2 z_z^2\right] z_{\rho
}^2+2 \left[9 \epsilon +(1+4\epsilon -5 \epsilon^2 ) z_z^2\right] z_{\rho }^4+\left[-1+2 (-5+\epsilon ) \epsilon \right] z_{\rho }^6\right\}}{2 z_z \left[1+(-1+\epsilon
) z_{\rho }^2\right]{}^2 \left\{(-1+\epsilon )^2 z_z^4-12 \epsilon  z_{\rho }^2+(1+2 \epsilon )^2 z_{\rho }^4+2 z_z^2 \left[6 \epsilon +(-1+\epsilon
) (1+2 \epsilon ) z_{\rho }^2\right]\right\}}\\  
&&\hspace*{8mm}
+A_2\, \big[ \left(\sqrt{1-\epsilon }-\sqrt{1+2 \epsilon }\right) \left(\left(-12
\epsilon +3 (-1+\epsilon ) \epsilon  z_z^2-(-1+\epsilon )^3 z_z^4\right) z_{\rho }^2+\left(3 \epsilon  (1+2 \epsilon )-\left(2-3 \epsilon +\epsilon
^3\right) z_z^2\right) z_{\rho }^4 \right. \nonumber\\
&&\hspace*{12mm}
\left.  +(-1+\epsilon )^2 (1+2 \epsilon ) z_{\rho }^6 + 12 \epsilon  z_z^2\right)\big]\big/
\big[2 \left(-1-2 \epsilon +\sqrt{1+\epsilon -2 \epsilon ^2}\right)
\left(1+(-1+\epsilon ) z_{\rho }^2\right){}^2 \left((-1+\epsilon )^2 z_z^4-12 \epsilon  z_{\rho }^2\right.\nonumber\\
&&\hspace*{12mm}
\left. +(1+2 \epsilon )^2 z_{\rho }^4+2 z_z^2 \left(6
\epsilon +(-1+\epsilon ) (1+2 \epsilon ) z_{\rho }^2\right)\right)\big]\nonumber\\
&&\hspace*{8mm}
+B_1\, \big[z_z \left(z_z^2 \left(-48 \epsilon +24 (-1+\epsilon ) \epsilon  z_z^2-(-1+\epsilon )^3 z_z^4\right) z_{\rho }^2+3
\left(8 \epsilon +8 \epsilon  (2+\epsilon ) z_z^2+(-1+\epsilon )^2 (-1+4 \epsilon ) z_z^4\right) z_{\rho }^4\right.\nonumber\\
&&\hspace*{12mm}
\left. +3 \left(-8 \epsilon  (1+2 \epsilon )+\left(1-9
\epsilon +8 \epsilon ^3\right) z_z^2\right) z_{\rho }^6+(-1+4 \epsilon  (3-2 (-3+\epsilon ) \epsilon )) z_{\rho }^8 + 24 \epsilon  z_z^4\right)\big]\big/
\big[\left(1+(-1+\epsilon) z_{\rho }^2\right) \nonumber\\
&&\hspace*{12mm}
\times  \left(z_z^2 \left(12 \epsilon +(-1+\epsilon )^2 z_z^2\right) z_{\rho }+2 \left(-6 \epsilon +(-1+\epsilon ) (1+2 \epsilon ) z_z^2\right)
z_{\rho }^3+(1+2 \epsilon )^2 z_{\rho }^5\right){}^2\big]\nonumber\\
&&\hspace*{8mm}
+B_2\, \big[(-1-2
\epsilon +\sqrt{1+\epsilon -2 \epsilon ^2}) \big((-1+\epsilon )^4 z_z^6 z_{\rho }^2+12 (1-7 \epsilon ) \epsilon  z_{\rho }^4+3 \epsilon
 (1-8 (-2+\epsilon ) \epsilon ) z_{\rho }^6-\left(1+\epsilon -2 \epsilon ^2\right)^2 z_{\rho }^8\nonumber\\
 &&\hspace*{12mm}
\left. +3 z_z^2 z_{\rho }^2 \left(8 \epsilon  (-1+4 \epsilon
)+2 (-1+\epsilon ) \epsilon  (1+8 \epsilon ) z_{\rho }^2+(-1+\epsilon )^2 (1+2 \epsilon ) z_{\rho }^4\right)\right. \nonumber\\
&&\hspace*{12mm}
+3 (-1+\epsilon ) z_z^4 \left(-4 \epsilon
+(-1+\epsilon ) z_{\rho }^2 \left(\epsilon +\left(-1+\epsilon ^2\right) z_{\rho }^2\right)\right)\big)\big]\big/
\big[(-1+\epsilon ) \left(\sqrt{1-\epsilon }-\sqrt{1+2
\epsilon }\right) \left(1+(-1+\epsilon ) z_{\rho }^2\right) \nonumber\\
&&\hspace*{12mm}
\times \left((-1+\epsilon )^2 z_z^4-12 \epsilon  z_{\rho }^2+(1+2 \epsilon )^2 z_{\rho }^4+2
z_z^2 \left(6 \epsilon +(-1+\epsilon ) (1+2 \epsilon ) z_{\rho }^2\right)\right){}^2\big] \nonumber\\
&&\hspace*{8mm}
+\big[\sqrt{1+2 \epsilon } (-1+4 \epsilon ) z_{\rho }^4-\sqrt{1+2
\epsilon } z_z^4 \left(1+5 \epsilon +3 (-1+\epsilon ) \epsilon  z_{\rho }^2\right)+\sqrt{1+2 \epsilon } z_z^2 z_{\rho }^2 \left(2+\epsilon -3 \epsilon
 (1+2 \epsilon ) z_{\rho }^2\right)\nonumber\\
&&\hspace*{12mm}
+z_z z_{\rho }^2 \left(-6 \epsilon +\left(1+\epsilon -2 \epsilon ^2\right) z_{\rho }^2\right)+z_z^3 \left(6 \epsilon
+(-2+\epsilon  (-2+13 \epsilon )) z_{\rho }^2\right)+ \left(1+\epsilon -2 \epsilon ^2\right) z_z^5\big]\big/\nonumber\\
&&\hspace*{8mm}
\big[\left(-1+(1+2 \epsilon ) z_z^2\right) \left(1+(-1+\epsilon ) z_{\rho }^2\right) \left((-1+\epsilon
)^2 z_z^4-12 \epsilon  z_{\rho }^2+(1+2 \epsilon )^2 z_{\rho }^4+2 z_z^2 \left(6 \epsilon +(-1+\epsilon ) (1+2 \epsilon ) z_{\rho }^2\right)\right)\big]\, .\nonumber
\end{eqnarray}
The disorder correction to the chemical potential, defined by Eq.~(\ref{eq:eqstgen}), is given by:
\begin{eqnarray}
\label{eq:appdelmu}
&&\hspace*{-3mm}\Delta_\mu =
A_1\, \big[
\left(24 \epsilon  z_{\rho }^2+(-2+\epsilon  (-17+10 \epsilon )) z_{\rho }^4+(-1+\epsilon )^2 (1+2 \epsilon ) z_{\rho }^6-(-1+\epsilon )^2 z_z^4
\left(2+(-1+\epsilon ) z_{\rho }^2\right)\right.\\
&&\hspace*{12mm}
+\left. z_z^2 \left(-24 \epsilon -(-1+\epsilon ) (4+17 \epsilon ) z_{\rho }^2-\left(2-3 \epsilon +\epsilon ^3\right)
z_{\rho }^4\right)\right)\big]\big/
\big[z_z \left(1+(-1+\epsilon ) z_{\rho }^2\right){}^2 \left((-1+\epsilon )^2 z_z^4-12 \epsilon  z_{\rho }^2\right.\nonumber\\
&&\hspace*{12mm}
\left. +(1+2 \epsilon
)^2 z_{\rho }^4+2 z_z^2 \left(6 \epsilon +(-1+\epsilon ) (1+2 \epsilon ) z_{\rho }^2\right)\right)\big]\nonumber\\
&&\hspace*{8mm}
-A_2\, \big[(-1+\epsilon ) \left(\sqrt{1-\epsilon }-\sqrt{1+2
\epsilon }\right)  \left(-30 \epsilon  z_{\rho }^2+(2+2 (13-5 \epsilon ) \epsilon ) z_{\rho }^4+(-1+\epsilon ) (1+2 \epsilon )^2 z_{\rho }^6\right.\nonumber\\
&&\hspace*{12mm}
\left. +(-1+\epsilon
)^2 z_z^4 \left(2+(-1+\epsilon ) z_{\rho }^2\right)+2 z_z^2 \left(15 \epsilon +(-1+\epsilon ) z_{\rho }^2 \left(2+13 \epsilon +(-1+\epsilon ) (1+2
\epsilon ) z_{\rho }^2\right)\right)\right)\big]\big/\nonumber\\
&&\hspace*{12mm}
\big[(-1-2 \epsilon +\sqrt{1+\epsilon -2 \epsilon ^2}) \left(1+(-1+\epsilon ) z_{\rho }^2\right){}^2
\left((-1+\epsilon )^2 z_z^4-12 \epsilon  z_{\rho }^2+(1+2 \epsilon )^2 z_{\rho }^4\right.\nonumber\\
&&\hspace*{12mm}
\left. +2 z_z^2 \left(6 \epsilon +(-1+\epsilon ) (1+2 \epsilon ) z_{\rho
}^2\right)\right)\big]\nonumber\\
&&\hspace*{8mm}
-B_1\, \big[2 z_z (36 \epsilon  (-1+3 \epsilon ) z_{\rho }^4+(2+\epsilon  (27+8 \epsilon  (-12+5 \epsilon ))) z_{\rho }^6-\left(1+\epsilon
-2 \epsilon ^2\right)^2 z_{\rho }^8+(-1+\epsilon )^3 z_z^6 \left(2+(-1+\epsilon ) z_{\rho }^2\right) \nonumber\\
&&\hspace*{12mm}
+3 z_z^2 z_{\rho }^2 \left(24 (1-2 \epsilon )
\epsilon +(-1+\epsilon ) z_{\rho }^2 \left(2+2 (9-16 \epsilon ) \epsilon +(-1+\epsilon ) (1+2 \epsilon ) z_{\rho }^2\right)\right)\nonumber\\
&&\hspace*{12mm}
+ 3 (-1+\epsilon
) z_z^4 \left(12 \epsilon +(-1+\epsilon ) z_{\rho }^2 \left(2+9 \epsilon +\left(-1+\epsilon ^2\right) z_{\rho }^2\right)\right))\big]\big/
\big[z_{\rho }^2
\left(1+(-1+\epsilon ) z_{\rho }^2\right) \left((-1+\epsilon )^2 z_z^4-12 \epsilon  z_{\rho }^2\right.\nonumber\\
&&\hspace*{12mm}
\left. +(1+2 \epsilon )^2 z_{\rho }^4+2 z_z^2 \left(6 \epsilon
+(-1+\epsilon ) (1+2 \epsilon ) z_{\rho }^2\right)\right){}^2\big]\nonumber\\
&&\hspace*{8mm}
+B_2\, \big[2 (-1-2 \epsilon +\sqrt{1+\epsilon -2 \epsilon ^2}) (-144
\epsilon ^2 z_{\rho }^2+6 \epsilon  \left(7+22 \epsilon -20 \epsilon ^2\right) z_{\rho }^4-2 (-1+\epsilon ) (1+2 \epsilon ) (-1+8 (-2+\epsilon )
\epsilon ) z_{\rho }^6\nonumber\\
&&\hspace*{12mm}
+3 (-1+\epsilon
)^2 z_z^4 \left(14 \epsilon +(-1+\epsilon ) z_{\rho }^2 \left(2+12 \epsilon +(-1+\epsilon ) (1+2 \epsilon ) z_{\rho }^2\right)\right)\nonumber\\
&&\hspace*{12mm}
+3 z_z^2 (48
\epsilon ^2+(-1+\epsilon ) z_{\rho }^2 (4 \epsilon  (7+11 \epsilon )+2 (-1+\epsilon ) (1+4 \epsilon  (3+2 \epsilon )) z_{\rho }^2+\left(1+\epsilon
-2 \epsilon ^2\right)^2 z_{\rho }^4))\nonumber\\
&&\hspace*{12mm}
+(-1+\epsilon )^2 (1+2 \epsilon )^3 z_{\rho }^8+(-1+\epsilon )^4 z_z^6 \left(2+(-1+\epsilon ) z_{\rho }^2\right))\big]\big/
\big[(-1+\epsilon ) \left(\sqrt{1-\epsilon }-\sqrt{1+2 \epsilon }\right) \nonumber\\
&&\hspace*{12mm}
\times \left(1+(-1+\epsilon
) z_{\rho }^2\right) \left((-1+\epsilon )^2 z_z^4-12 \epsilon  z_{\rho }^2+(1+2 \epsilon )^2 z_{\rho }^4+2 z_z^2 \left(6 \epsilon +(-1+\epsilon )
(1+2 \epsilon ) z_{\rho }^2\right)\right){}^2\big]\nonumber\\
&&\hspace*{8mm}
+\big[2 ((-1+\epsilon ) (1+2 \epsilon )^{3/2} z_z^4+(-1+\epsilon )^2 (1+2 \epsilon ) z_z^5+\sqrt{1+2 \epsilon } z_{\rho }^2 \left(6 \epsilon
+(-1+\epsilon ) (1+2 \epsilon ) z_{\rho }^2\right)\nonumber\\
&&\hspace*{12mm}
+z_z^3 \left(9 \epsilon  (1+\epsilon )+(-1+\epsilon ) (2+\epsilon ) (1+2 \epsilon ) z_{\rho }^2\right)+\sqrt{1+2
\epsilon } z_z^2 \left(-6 \epsilon +(2+(2-13 \epsilon ) \epsilon ) z_{\rho }^2\right)\nonumber\\
&&\hspace*{12mm}
+z_z z_{\rho }^2 \left(-9 \epsilon +\left(1+3 \epsilon -4 \epsilon
^3\right) z_{\rho }^2\right))\big]\big/
\big[\left(-1+(1+2 \epsilon ) z_z^2\right) \left(1+(-1+\epsilon ) z_{\rho }^2\right) \nonumber\\
&&\hspace*{12mm}
\times \left((-1+\epsilon )^2 z_z^4-12
\epsilon  z_{\rho }^2+(1+2 \epsilon )^2 z_{\rho }^4+2 z_z^2 \left(6 \epsilon +(-1+\epsilon ) (1+2 \epsilon ) z_{\rho }^2\right)\right)\big]\nonumber\\
&&\hspace*{8mm}
+\frac{8}{z_z+z_{\rho }}\, T\left(\frac{-z_z+z_{\rho }}{z_z+z_{\rho }}\right)\, .\nonumber
\end{eqnarray}
\end{widetext}

\end{document}